\documentclass[fleqn,usenatbib]{mnras}

\usepackage{newtxtext,newtxmath}

\usepackage[T1]{fontenc}
\usepackage{ae,aecompl}
\usepackage{placeins}

\usepackage{caption}
\usepackage[flushleft]{threeparttable}
\usepackage{pifont}
\usepackage{xcolor}

\usepackage{graphicx}	
\usepackage{amsmath}	
\usepackage{amssymb}	
\usepackage{wrapfig}
\usepackage{supertabular,booktabs}

\title[Source-tailored versus generic modelling]{Physicochemical models: source-tailored or generic?}

\author[B. M. Kulterer et al.]{
Beatrice M. Kulterer,$^{1}$\thanks{E-mail: beatrice.kulterer@csh.unibe.ch}
Maria N. Drozdovskaya,$^{1}$
Audrey Coutens,$^{2}$
\newauthor
S\'{e}bastien Manigand,$^{3}$
Gwendoline St\'{e}phan$^{4}$
\\
$^{1}$Center for Space and Habitability, University of Bern, Gesellschaftsstrasse 6, 3012 Bern, Switzerland\\
$^{2}$Laboratoire  d'astrophysique de Bordeaux, Univ. Bordeaux, CNRS, B18N, all\'{e}e Geoffroy Saint-Hilaire, 33615 Pessac, France\\
$^{3}$Center for Star and Planet Formation, Niels Bohr Institute \& Natural History Museum of Denmark, University of Copenhagen, \O ster Voldgade 5-7, \\
1350 Copenhagen K., Denmark\\
$^{4}$ Department of Chemistry, University of Virginia, Charlottesville, VA 22904, USA
}

\date{Accepted 2020 August 10. Received 2020 August 10; in original form 2020 May 22}

\pubyear{2020}

\begin{document}
\label{firstpage}
\pagerange{\pageref{firstpage}--\pageref{lastpage}}
\maketitle

\begin{abstract}
Physicochemical models can be powerful tools to trace the chemical evolution of a protostellar system and allow to constrain its physical conditions at formation. The aim of this work is to assess whether source-tailored modelling is needed to explain the observed molecular abundances around young, low-mass protostars or if, and to what extent, generic models can improve our understanding of the chemistry in the earliest stages of star formation. The physical conditions and the abundances of simple, most abundant molecules based on three models are compared. After establishing the discrepancies between the calculated chemical output, the calculations are redone with the same chemical model for all three sets of physical input parameters. With the differences arising from the chemical models eliminated, the output is compared based on the influence of the physical model. Results suggest that the impact of the chemical model is small compared to the influence of the physical conditions, with considered timescales having the most drastic effect. Source-tailored models may be simpler by design; however, likely do not sufficiently constrain the physical and chemical parameters within the global picture of star-forming regions. Generic models with more comprehensive physics may not provide the optimal match to observations of a particular protostellar system, but allow a source to be studied in perspective of other star-forming regions.

\end{abstract}

\begin{keywords}
ISM: abundances -- astrochemistry -- protoplanetary discs -- stars: protostars 
\end{keywords}

\section{Introduction}
Cold, dense cores may be starless, prestellar, or protostellar in nature. Initially, they are all characterized by a spatial extent in the range of $\sim$~0.1 pc, temperatures of $\sim$~10~K, and typical densities of a few 10$^4$~cm$^{-3}$ (\citealt{1989ApJS...71...89B,2007ARA&A..45..339B}). With time these slowly rotating cores with a typical rotation rate ($\Omega$) of $\sim$~1~km~s$^{-1}$~pc$^{-1}$ (\citealt{1987ARA&A..25...23S,1993ApJ...406..528G}) can concentrate mass If their central density increases above 10$^5$~cm$^{-3}$, they are deemed to be prestellar and display signs of kinematic and chemical evolution (e.g., \citealt{2005ApJ...619..379C,2008ApJ...683..238K}). Once they become unstable and collapse in an inside-out manner, a central protostar accompanied by a disc is formed (\citealt{1987ARA&A..25...23S}). Both are enshrouded by an infalling envelope of dust and gas, which accretes onto the disc and the protostar. If the formation and destruction rates of molecules are quantified, they can function as diagnostic tools of the evolving physical structures of these systems, because the rates of the chemical processes depend on the physical conditions and time. Comparing observations to predicted model abundances gives insights into the history of such systems. 
\newline 
\noindent Since the first detection of an interstellar molecule in the 1930s (e.g., \citealt{1937ApJ....86..483S}), more than 200 have been detected in the interstellar medium (ISM; \citealt{2018ApJS..239...17M}). While the majority of the detected species consists of two or three atoms, complex organic molecules (COMs) have been detected in prestellar cores and around protostars of all masses (e.g., \citealt{2019A&A...625A.147A}, \citealt{2019A&A...631A.142G}). Following the definition given by \cite{2009ARA&A..47..427H}, a COM consists of at least six atoms and contains the element carbon. The simplest in structure is methanol (CH$_3$OH), which is frequently detected in the ISM (e.g., \citealt{2016A&A...595L...4L}, \citealt{2018ApJS..236...45G}, \citealt{2019A&A...622A.141C}). Its formation occurs on dust grain surfaces at dust temperatures of $\sim$~12~K by subsequent hydrogenation of CO, which has been quantified by theoretical studies and in the laboratory (e.g., \citealt{1982A&A...114..245T}, \citealt{2009A&A...505..629F}). \cite{2009A&A...504..891O} showed in laboratory experiments that radicals, produced by the photodissociation of UV-irradiated (ultraviolet) methanol ices, proceed to form species with a higher degree of complexity, e.g., glycolaldehyde (CH$_2$OHCHO). Moreover, laboratory studies produced species as complex as glycerol (HOCH$_2$CH(OH)CH$_2$OH) starting from the hydrogenation of CO (\citealt{2017ApJ...842...52F}). Thus, a variety of the observed complex species is believed to form on icy interstellar grains (\citealt{2008PrSS...83..439W}). 
\newline 
\noindent Astrochemical models provide a powerful tool to link observed molecules in interstellar space to theoretical and experimental findings. Physical parameters (e.g., gas density, gas and dust temperatures, UV flux) are derived from observational data, which are then analyzed following radiative transfer theory. The derived physical parameters can then be used to tailor a model specifically towards the observed source or serve as input for a model that aims to provide a more generic picture of a typical source of its type. Models can target a single structure of a system, a certain time during the system's evolution, or give its full evolutionary history. For instance, dedicated models have been developed for protoplanetary discs (e.g., \citealt{2014A&A...563A..33W}, \citealt{2018ApJ...856...85S}, \citealt{2019MNRAS.484.1563W}). Others investigated the chemical evolution of pre- and protostellar systems (e.g., \citealt{2014ApJ...791....1T}, \citealt{2017ApJ...842...33V}) or connected astrochemical modelling to planet formation models to investigate the composition of planetary embryos (e.g., \citealt{2019A&A...627A.127C}). Input for the chemical networks used in models is gained from laboratory experiments, where several processes are considered over a range of physical conditions for a selection of molecules, including processes in the gas phase and in ices, as well as theoretical computations. Physicochemical models combine these results alongside data taken from observational studies to predict the molecular abundances in interstellar environments. 
This task is challenging due to missing measurements of many reaction rates and a lack of clarity on crucial parameters in processes such as diffusion, desorption, and reaction barriers (e.g., \citealt{2017SSRv..212....1C}, \citealt{2017ApJ...844...71P}, \citealt{2017MolAs...9....1W}). Moreover, physicochemical models are constrained by computational limits; not all details can be considered. Hence, different models are tailored to focus on selected physical or chemical processes, while making simplified assumptions on other parts.
\newline 
\noindent In this work, three physicochemical models are analyzed.
These models have already been published in \cite{2014MNRAS.445.2854W} for model W14-M20 (based on \citealt{2008ApJ...674..984A}, last published in \cite{2020arXiv200704000M}), \cite{2016MNRAS.462..977D} for model D16 and St\'{e}phan et al. (2020, in prep.) for model S20, and will be referred to as such in the text. Two out of the three models, W14-M20 and S20, are 1D and constructed specifically for IRAS 16293-2422 (hereafter, IRAS 16293), while D16 represents a more generic, 2D approach to the formation of a low-mass protostar. Located in $\rho$~Ophiuchus, the system is composed of two sources, A and B. Recently, \cite{2020arXiv200511954M} have shown that source A is itself a close binary. The B component that is targeted by the 1D models has an estimated mass of 0.1~M$_{\odot}$ (\citealt{2018A&A...612A..72J}). Sources A and B were targeted by the Protostellar Interferometric Line Survey (PILS; \cite{2016A&A...595A.117J}) executed with ALMA (Atacama Large Millimeter Array) and have shown to exhibit a variety of molecules, including COMs (e.g., \citealt{2017MNRAS.469.2219L}, \citealt{2018MNRAS.476.4949D}, \citealt{2018A&A...620A.170J}, \citealt{2019A&A...631A.137C}, \citealt{2019A&A...623A..69M}). 
\newline 
\noindent This work assesses whether source-tailored modelling is needed to explain the observed molecular abundances around young stellar objects or if, and to what extent, generic models can improve our understanding of the chemistry in the earliest stages of star formation. The physical conditions and the abundances of simple, most abundant molecules based on these three physicochemical models are compared. These models consider and target different components of the evolving protostellar system and use different methods to calculate the physical parameters. After establishing the discrepancies between the calculated chemical outputs, the calculations are redone with the same chemical model for all three sets of physical input parameters. This eliminates the differences arising from the chemical networks and gives direct access to the differences stemming solely from the physical models. 
\newline 
\noindent Section \ref{Models} gives a summary of the physical and the chemical models. Section \ref{Results} presents the comparison of the physical models and describes the obtained molecular abundances with the respective chemical codes and then with the same model, followed by the discussion in Section \ref{Discussion}. The conclusions are raised in Section \ref{Conclusion}.

\section{Models}\label{Models}
Here, a short summary of the physical and chemical models used to obtain the results is given. All models assume T$_{\rm dust}$~=~T$_{\rm gas}$. 

\subsection{Physical models}

\begin{table}
    \centering
     \caption{Physical precollapse parameters for the individual models.}
    \begin{tabular}{|l|l|l|l|}
    \hline 
     Parameter    & W14-M20$^{\rm a}$ & S20$^{\rm b}$ & D16$^{\rm c}$ \\
     \hline
     T$_{\rm dust,gas}$~[K]  & 10  & 8  & 10 \\
     n$_{\rm H}$~[cm$^{-3}$] & 2.0~$\times$~10$^4$  & 2.3~$\times$~10$^4$  & 4.0~$\times$~10$^4$ \\
     A$_{\rm v}$~[mag] & 4.5 & 5.0-7.5 & 10 \\
     $\zeta _{\rm CR}$~[s$^{-1}$] & 1.3~$\times$~10$^{-17}$  & 1.3 $\times$~10$^{-17}$ & 5.0~$\times$ 10$^{-17}$\\
     t$_{\rm precollapse}$~[yr] & 1.0~$\times$~10$^6$  & 1.0~$\times$~10$^6$  & 3.0~$\times$~10$^5$  \\
      t$_{\rm collapse}$~[yr] & 3.4~$\times$~10$^5$  & 3.2~$\times$~10$^5$  & 2.5~$\times$~10$^5$  \\
     t$_{\rm *birth}$ [yr] & 2.5~$\times$~10$^5$ & 3.1~$\times$~10$^5$ & 2.0~$\times$~10$^4$ \\
     t$_{\rm *birth}$ [t$_{\rm  collapse}$] & 0.73 & 0.97 & 0.08 \\
    
     \hline 
    \end{tabular}
    \begin{tablenotes}
    \item[*] $^{\rm a}$ physical model from \cite{2014MNRAS.445.2854W}, based on \cite{2008ApJ...674..984A}, last published in \cite{2020arXiv200704000M}; $^{\rm b}$  St\'{e}phan et al. (2020, in prep.); $^{\rm c}$ last published in \cite{2016MNRAS.462..977D}
    \end{tablenotes}
    \label{initial_parameters_physics}
\end{table}

\subsubsection{W14-M20}\label{W14-M20} 
This 1D model follows the evolution of a hydrostatic, prestellar core (\citealt{2000ApJ...531..350M}) to a protostellar object under the assumption of free-fall collapse and has been used to model the envelope of IRAS 16293 (\citealt{2008ApJ...674..984A}, most recently updated in \citealt{2014MNRAS.445.2854W}). The prestellar core is characterized by an atomic H density (n$_{\rm H}$) of 2.0~$\times$~10$^4$~cm$^{-3}$, a visual extinction (A$_{\rm V}$) of 4.5 mag, a cosmic-ray ionisation rate ($\zeta _{\rm CR}$) of 1.3~$\times$ 10$^{-17}$~s$^{-1}$, and gas and dust temperatures (T$_{\rm dust}$) of 10~K (see Table \ref{initial_parameters_physics}). These parameters are kept constant for t$_{\rm precollapse}^{\rm W14-M20}$ = 1.0~$\times$~10$^6$~yr and then are evolved for t$_{\rm collapse}^{\rm W14-M20}$. Compressional heating leads to an accumulation of a central density, and 560 yr before the protostellar birth, the first hydrostatic core (FHC) is formed. Only recently a candidate FHC has been observed \cite{2020ApJ...890..129K}. Theory suggests FHC formation occurs as soon as the central density is high enough ($\sim$~10$^{13}$~g~cm$^{-3}$) for the inner region to become opaque to radiation (\citealt{2014prpl.conf..195D}). The FHC is characterized by a radius of 1~au in this model. When its density increases to 10$^{7}$~g~cm$^{-3}$ and the temperature to 2~000~K, a protostar is born at 2.5~$\times$~10$^5$~yr into the collapse (=~t$_{\rm *birth}^{\rm W14-M20}$ in Table \ref{initial_parameters_physics}). Thereafter, the model follows the evolution for an additional 9.3~$\times$~10$^4$~yr. As the observed densities obtained from single-dish multi-wavelength dust and molecular observations of the envelope of the IRAS 16293-2422 system by \cite{2010A&A...519A..65C} are about 10 times larger than the ones given by the model, the calculated densities are multiplied by this factor at all times and radii (\citealt{2014MNRAS.445.2854W}, \citealt{2018MNRAS.481.5651A}, \citealt{2019A&A...623L..13C}, \citealt{2020arXiv200704000M}). Model W14-M20 assumes that the core is embedded in an ambient cloud. As a consequence, the visual extinction of the prestellar phase is increased by three magnitudes to the given value of 4.5 mag (\citealt{2008ApJ...674..984A}). During collapse, the model by \cite{2008ApJ...674..984A} calculates the visual extinction via the column density of hydrogen nuclei (N$_{\rm H}$) from the outer core edge to the position of each parcel via A$_{\rm V}$~=~N$_{\rm H}$/(1.59~$\times$~10$^{21}$~cm$^{-2}$)~mag. Thus, the attenuating column of material is that from the outer envelope shell to the position of a parcel at time $t$. No additional radiative transfer solver is used for the calculation of the radiation field nor the temperature distribution. The temperature is computed parametrically according to the model of \cite{2000ApJ...531..350M}. This model considers an external UV field, which includes surrounding stars and CRs that produce FUV photons through excitation of H$_2$. The model only considers sufficiently dense parts of the envelope, where the influence of the internal UV flux from the emerging protostar is assumed to be negligible. Three representative trajectories, that end up at distances of 15, 62.4 and 125~au from the protostar, are considered in this work to trace the physical evolution of collapsing material. 

\subsubsection{S20}\label{S20}
The S20 model is presented in St\'{e}phan et al. (in prep.) and is tailored towards IRAS 16293B. The initial physical conditions of the precollapse phase are n$_{\rm H}$ = 2.3~$\times$~10$^4$~cm$^{-3}$, T$_{\rm dust}$ = 8~K, A$_{\rm V}$ = 5.0 \textendash 7.5~mag (depending on the radial distance from the protostar), and $\zeta _{\rm CR}$ = 1.3~$\times$ 10$^{-17}$~s$^{-1}$ and are evolved for t$_{\rm precollapse}^{\rm S20}$~=~1.0~$\times$~10$^6$~yr (Table \ref{initial_parameters_physics}). Assuming free-fall collapse, the collapsing envelope is traced by tracking the position of gas parcels for 3.2~$\times$~10$^5$~yr, the considered duration of the collapse. The density profile is derived from observations by \cite{2002A&A...390.1001S} and is assumed to equal the density profile at the end of the collapse. This is used to calculate the power law density profiles of the individual trajectories. The derived parameters of the best fit model in \cite{2002A&A...390.1001S} set the radius of the inner envelope to 32~au, which equals the final position of the innermost trajectory. The accretion model for the protostar is taken from \cite{2009ApJ...691..823H}, where the accretion rate (\.M$_{*}$) is kept at 10$^{-5}$~M$_{\odot}$~yr$^{-1}$ during the collapse. The luminosity of the central protostar (L$_{*}$) is assumed to be 10.5~L$_{\odot}$ based on the protostellar accretion model of \cite{2009ApJ...691..823H}, which turns on at t$_{\rm *birth}^{\rm S20}$~=~3.1~$\times$~10$^5$~yr. Under the assumption that T$_{\rm gas}$~=~T$_{\rm dust}$ at all times, the dust temperature profiles of the trajectories are calculated with the radiative transfer code \textsc{Radmc-3d} (\citealt{2012ascl.soft02015D}). The F$_{\rm UV}$ field is set to the arbitrary value of 10$^{-8}$~G$_0$, where G$_0$ is 1.6~$\times$~10$^{-3}$~erg~cm$^{-2}$~s$^{-1}$, at all times throughout the collapse. This work considers trajectories that end at distances of 32, 49.7, 61.9, 101, 125, 163, 203, and 232~au.

\subsubsection{D16}\label{D16}
The D16 model assumes an axisymmetric, semi-analytic collapse in 2D and tracks the physical evolution of the system including disc formation. This dynamic collapse model is adapted from \cite{1977ApJ...214..488S} and further developed by \cite{2009A&A...495..881V},  \cite{2010A&A...519A..28V}, \cite{2011A&A...534A.132V}, and \cite{2013A&A...555A..45H}. Starting with the precollapse parameters T$_{\rm dust}$ = 10~K, n$_{\rm H}$~=~4.0~$\times$~10$^4$~cm$^{\rm -3}$, A$_{\rm V}$~=~10~mag, and $\zeta _{\rm CR}$~=~5.0~$\times$~10$^{-17}$~s$^{-1}$ that are evolved for t$_{\rm precollapse}^{\rm D16}$~=~3.0~$\times$~10$^5$~yr, the collapse proceeds for 2.5~$\times$~10$^5$~yr (Table \ref{initial_parameters_physics}). Adapted from \cite{2005ApJ...627..293Y}, the model forms its FHC during the first 2.0~$\times$~10$^4$~yr of the collapse. The radius of the FHC  is estimated to be 5~au (\citealt{2000ApJ...531..350M}) in this model. As the collapse continues, it transitions down to its initial protostellar radius of $\sim$~2.5~R$_{\odot}$ following the calculations of \cite{1991ApJ...375..288P} within <~100~yr independent of other parameters (\citealt{2009A&A...495..881V}). Hence, at 2.0~$\times$~10$^4$~yr, R$_{\rm *}$ equals the radius calculated by \cite{1991ApJ...375..288P}. As soon as the decrease of the radius stops: the protostar is born. At early times, the protostellar luminosity is driven by shock accretion. At later times, in the pre-main sequence phase of stellar evolution, the luminosity stems from gravitational contraction and deuterium burning (based on \citealt{1994ApJS...90..467D,2009A&A...495..881V}). The now pre-main sequence star continues to grow by accreting mass from the forming disc and the infalling envelope. The model is evolved until it reaches its so-called accretion time (t$_{\rm acc}$, for this work: t$_{\rm collapse}$), which is defined as the end of the primary accretion phase onto the star. At this point, the outer shell of the envelope has traveled inwards to reach the protoplanetary disc. This depends on the initial parameters that are chosen to solve the hydrodynamics equations of a collapsing isothermal sphere (\citealt{1977ApJ...214..488S}). The parameters include the initial core mass, which is set to 1~M$_{\odot}$, the gravitational constant (G), the effective sound speed (c$_{\rm s}$) and a constant m$_0$~=~0.975, that stems from the analytical solution of the collapse model (\citealt{1977ApJ...214..488S}). Thus, this model traces the evolution after the stellar birth for 2.3~$\times$~10$^5$~yr. At all times during this evolution, 2D density and velocity distributions are obtained. To compute the temperatures (T$_{\rm dust}$~=~T$_{\rm gas}$) and the stellar radiation field, the results of the collapse are fed into the radiative transfer code \textsc{Radmc-3d} (\citealt{2012ascl.soft02015D}). This model does not account for sources of external UV, because it assumes that the star-forming system is deeply embedded in a core. However, it includes the protostar after t$_{\rm *birth}$ as a source of internal UV and takes CRs, which produce FUV photons through excitation of H$_2$, into account. The resulting F$_{\rm UV}$ flux as calculated by \textsc{Radmc-3d} can be converted to the visual extinction (A$_{\rm V}$) via the scaling relation:
\begin{equation}\label{eq_AV}
    A_{\rm V} = \tau_{\rm UV, eff} / 3.02,
\end{equation}
where $\tau_{\rm UV,eff}$ is the effective UV extinction (\citealt{1978ApJ...224..132B}) calculated with:
\begin{equation}\label{eq_tau}
    \tau_{\rm UV, eff} = -\rm ln \left( \frac{F_{\rm UV}}{\pi \times \int_{F_{UV}} B _\lambda~(T_*)~d\lambda~\times~R_*^2 /(R^2 + z^2)} \right).
\end{equation}
The denominator equals the blackbody radiation over the F$_{\rm UV}$ wavelength range from  912 to 2066 \r{A} (corresponding to 6.0-13.6~eV) with geometrical dilution, $\pi$ accounts for the radiation stemming from one hemisphere towards a point in the envelope (\citealt{2015MNRAS.451.3836D}). The stellar radius is denoted as R$_{*}$, R is the radial distance from the protostar, z describes the scale height. Here, the column of attenuating material is that between a position $(R,z)$ at time $t$ and the center of the system, which is where the radiating emerging protostar is located.
\newline
\noindent Two disc cases are discussed in detail in \cite{2014MNRAS.445..913D} and \cite{2016MNRAS.462..977D}. For this work, the case denoted as ``infall-dominated disc", or also ``case 7", is used. As can be seen in Fig. 5 of \cite{2014MNRAS.445..913D}, this model results in an extended, massive disc (R$_{\rm disc}$ $\sim$~300~au and M$_{\rm disc}$~$\sim$~0.44~M$_{\odot}$ at t$_{\rm collapse}^{\rm D16}$) with densities up to $\sim$~10$^{12}$~cm$^{-3}$ in the midplane in the proximity of the protostar. The resulting disc is cold, with dust temperatures ranging between $\sim$~20 and 100~K. Dust temperatures $>$~150~K are obtained in the outflow cavities and the most inner regions. The protostellar mass at t$_{\rm collapse}^{\rm D16}$ equals $\sim$~0.56~M$_{\odot}$.
For this work, the trajectories at 10.8, 20.6, 30.2, 40.2, and 46.7~au from \cite{2016MNRAS.462..977D} are investigated.

\subsection{Chemical models}\label{section_chemicalmodels}
In this section, a brief description of the chemical models that are used to calculate the molecular abundances is given. For detailed discussions of the included mechanisms the reader is referred to the respective publications: \cite{2016MNRAS.459.3756R} for W14-M20, \cite{2013ApJ...765...60G} for S20, and \cite{2014A&A...563A..33W} for D16. Table \ref{reactions} lists the considered reaction mechanisms of these chemical codes; the individual subsections describe their specific details. 
\newline 
\noindent In the three chemical models studied in this work, gaseous and solid phases are considered. The solid phase corresponds to the icy mantles that cover the dust grains. In the case of three-phase models (S20 and W14-M20), the icy mantle is further partitioned into a bulk and a surface layer. In the subsequent Sections 3.2~-~3.5, the three physicochemical models will be compared in terms of simple, most abundant molecules. The chosen species represent molecules common to the majority of existing chemical codes of pre- and protostellar cores that are frequently observed in different environments of the ISM. The nine molecules considered are carbon monoxide (CO), carbon dioxide (CO$_2$), water (H$_2$O), formaldehyde (H$_2$CO), methanol (CH$_3$OH), methane (CH$_4$), ammonia (NH$_3$), molecular nitrogen (N$_2$), and hydrogen sulfide (H$_2$S). In this work, atomic abundances are given relative to n$_{\rm H}$ and molecular abundances relative to n$_{\rm H_{2}}$.  
\newline 
\noindent In all physicochemical models, the chemistry is calculated in two steps. First, under constant precollapse physical parameters (Table \ref{initial_parameters_physics}) the initial atomic abundances (Table \ref{initial_abundances}) are evolved for t$_{\rm precollapse}$. The obtained molecular abundances (Table \ref{initial_molecular}) are then used as initial input for the second step: the modelling of the collapse phase under its changing physical conditions.

\begin{table}
    \centering
    \caption{Considered reaction mechanisms in the three chemical codes.}
    \begin{tabular}{|l|l|}
    \hline 
    Chemical phase     &  Reaction mechanism \\
    \hline
    Gas phase & 2-body associations (typical: neutral- \\
    & neutral reactions, ion-molecule reactions) \\
    & photodissociation$^{\rm a}$ \\
    & direct cosmic ray (CR) ionisation \\
    \hline 
    Gas-grain interactions & thermal desorption \\
    & non-thermal desorption: \\
    & \hspace{0.5cm} photodesorption$^{\rm a}$ \\
    & \hspace{0.5cm} reactive desorption  \\
    & spot-heating by CRs  \\
    & adsorption \\
    \hline
    Solid phase & 2-body associations (typical: radical- \\
    & radical associations via thermal \\
    & hopping and/or quantum tunneling) \\
    & photodissociation$^{\rm a}$  \\
    \hline 
    \end{tabular}
    \begin{tablenotes}
    \item[*] $^{\rm a}$ includes all sources of UV photons included in the specific model: stellar and interstellar UV photons, internally generated UV photons produced by the de-excitation via fluorescence cascades of H$_2$ molecules excited by CR impacts
    \end{tablenotes}
    \label{reactions}
\end{table}

\subsubsection{\textsc{NAUTILUS}}
\textsc{Nautilus} is a three-phase chemical code used with W14-M20. The bulk and the surface vary in chemical reactivity. The two outermost ice monolayers correspond to the surface. Exchange between the bulk and the surface layers occurs via swapping, which describes the transport of each individual species from the bulk to the surface and vice versa; the net transfer rate equals 0. The swapping rate from the bulk to the surface depends on the amount of layers in the bulk.
Thermal and non-thermal desorption, and accretion are considered only for the surface. Photodissociation and diffusion are taken into account for the surface and the bulk, but the diffusion rates for the bulk species are kept smaller: the diffusion barrier for the surface is set to E$_{\rm diff}^{\rm s}$~=~0.4~$\times$~E$_{\rm des}$ (desorption energy), while for the bulk E$_{\rm diff}^{\rm b}$~=~0.8~$\times$~E$_{\rm des}$ is used. Recent dedicated theoretical efforts of \cite{2019ApJ...876..140S} corroborate the lack of bulk diffusion at low temperatures. Moreover, solid-phase reactions are allowed in both ice components. No direct interactions between the gas phase and the bulk are considered. The Langmuir-Hinshelwood mechanism is the sole grain-surface reaction mechanism that is used (\citealt{2016MNRAS.459.3756R}). The rate equation approach for gas interactions and grain-surface chemistry is adapted from \cite{1992ApJS...82..167H} and \cite{1993MNRAS.261...83H}. The chemical network of \cite{2016MNRAS.459.3756R} has recently been extended by \cite{2020arXiv200704000M} to include larger COMs.

\subsubsection{\textsc{MAGICKAL}}

To calculate the molecular abundances with the S20 model, the three-phase model \textsc{Magickal} (Model for Astrophysical Gas and Ice Chemical Kinetics and Layering; \citealt{2013ApJ...765...60G}) was used. \textsc{Magickal} uses also the rate equation approach (\citealt{1992ApJS...82..167H}, \citealt{1993MNRAS.261...83H}) for gas and grain-surface reactions; and considers active chemistry in the gas phase, the icy surface of the dust particles, and the bulk ice. COM formation up to glycine is considered. In total, the chemical network includes 1369 species and over 21~000 reactions. As for \textsc{Nautilus}, the swapping mechanism between the surface and the bulk produces no net transfer between the two phases. The swapping barrier is set to E$_{\rm swap}$~=~0.7~$\times$~E$_{\rm des}$. For the most important species, individual $x$ in E$_{\rm diff}$~=~$x$~$\times$~E$_{\rm des}$ are tabulated (Table 4 in \citealt{2013ApJ...765...60G}). As for \textsc{Nautilus}, accretion from the gas onto the surface, and thermal and non-thermal desorption from the surface are allowed. No direct reactions between the bulk and the gas take place; the bulk material must first be transferred to the surface. Solid-phase reactions, diffusion, and photodissociation are allowed for the surface and the bulk. Reaction-diffusion competition is accounted for via activation-energy barriers. Quantum tunneling is allowed and the Languir-Hinshelwood is the only grain-surface reaction mechanism (\citealt{2013ApJ...765...60G}). The tunneling process is adapted from \cite{1992ApJS...82..167H}, where it is parameterised by an expression for tunneling through a rectangular potential. The barrier width is estimated from calculations (\citealt{2011ApJ...735...15G}).

\subsubsection{D16}
Details about the chemical code in the D16 model can be found in \cite{2014MNRAS.445..913D}, \cite{2014A&A...563A..33W}, \cite{2016MNRAS.462..977D}, and the references therein. Contrary to the other two models, this model treats the bulk and the surface as one equally chemically active  phase (i.e., a two-phase model). The included gas-phase network is the RATE12 release of the UMIST Database for Astrochemistry (UDfA; \citealt{2013A&A...550A..36M}). The rate equation approach by \cite{1992ApJS...82..167H} and \cite{1993MNRAS.261...83H} is adapted for gas and grain-surface reactions, as well as their quantum tunneling parameterisation. Again, quantum tunneling is allowed, the Langmuir-Hinshelwood mechanism is the sole mechanism for grain-surface chemistry (\citealt{2014A&A...563A..33W}).  Photodissociation and photoionization induced by the UV field are included in the network. Multiple families of COMs are included in the network. For all reactions E$_{\rm diff}$ is set to $0.3$~$\times$~E$_{\rm des}$. A coverage factor included in the calculation of the photodesorption rate assures that photodesorption occurs only in the top two monolayers of the surface, as suggested by experiments (\citealt{2012PCCP...14.9929B}).

\section{Results}\label{Results}

\subsection{Physical parameters}\label{Results_physics_text}
\begin{figure*}
	\includegraphics[width=2.0\columnwidth]{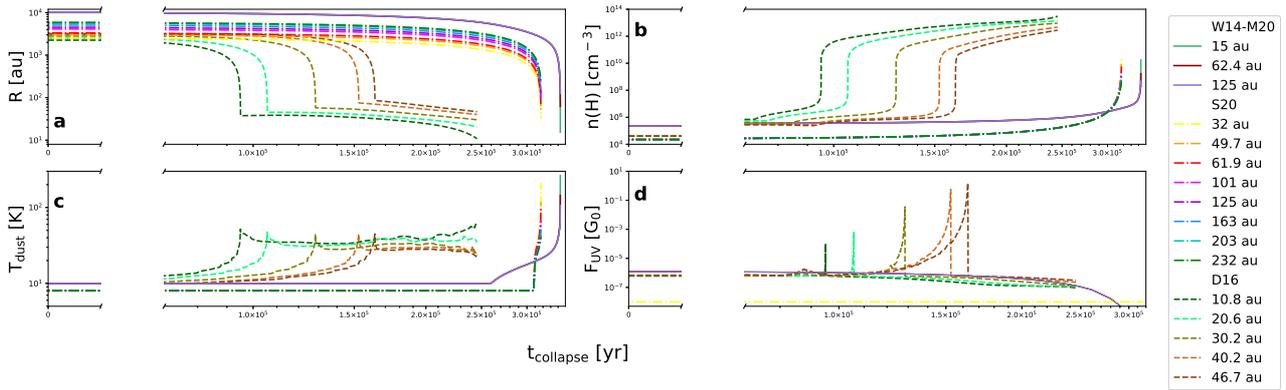}
    \caption{Evolution of the physical parameters during the collapse. Panel \ref{physics_results}a (upper left) depicts the radius in au, \ref{physics_results}b (upper right) the gas density in cm$^{-3}$, \ref{physics_results}c (lower left) the dust temperature in K, and \ref{physics_results}d (lower right) the F$_{\rm UV}$ flux relative to G$_0$. The x-axis gives the collapse time in years. Solid lines describe W14-M20, dotted-dashed S20, and dashed D16 with the colors corresponding to different final positions of the trajectories.}
    \label{physics_results}
\end{figure*}

Physical parameters as calculated by the three physical models are depicted in Fig. \ref{physics_results}.
While discussing the different physical parameters the inequality of the timescales should be kept in mind. In the case of D16, no time steps after 2.5~$\times$~10$^5$~yr are considered, as the accretion time has elapsed by this point. In the other two models, the protostars have not been born yet by this time. This occurs only after 2.5~$\times$~10$^5$~yr and 3.1~$\times$~10$^5$~yr for W14-M20 and S20, respectively, with the evolution being traced for 3.4~$\times$~10$^5$~yr and 3.2~$\times$~10$^5$~yr in total, respectively.

\subsubsection{Radial position}
Panel \ref{physics_results}a traces the distance to the central source during the collapse. All trajectories emerge from far out in the envelope; their infall paths start between $\sim$~2~000 - 6~000~au for S20, $\sim$~2~000 - 3~000~au for D16, and $\sim$~10~000 au for W14-M20. The infall pathways show that the material remains in the outer envelope at distances beyond 1~000 au for $\sim$~2.8~$\times$~10$^5$~yr ($\sim$~0.88~t$_{\rm collapse}^{\rm S20}$) for S20 and 3.4~$\times$~10$^5$~yr ($\sim$~0.99~t$_{\rm collapse}^{\rm W14-M20}$) for W14-M20. For S20, the innermost trajectory ends at the assumed inner envelope boundary at 32~au (\citealt{2002A&A...390.1001S}). Even at such proximity to the protostar this model never probes the material of the protoplanetary disc, because a disc is not considered. For W14-M20, the innermost trajectory at 15~au should also be in the protoplanetary disc regime, but the disc is not modeled in this case either. In the case of D16, the innermost trajectory breaks into the inner 100~au, which corresponds to the disc in this model, after $\sim$~8.0~$\times$~10$^4$~yr (0.33~t$_{\rm collapse}^{\rm D16}$). All considered trajectories follow this behaviour, the outermost trajectory reaches the innermost 100~au after $\sim$~1.6~$\times$~10$^5$~yr (0.66~t$_{\rm collapse}^{\rm D16}$). The 2D approach of this model also gives information about the scale height (z). At the end of the collapse, the considered trajectories reside in the midplane at z~$\sim$~0.01~au (\citealt{2016MNRAS.462..977D}). 

\subsubsection{Density}
Panel \ref{physics_results}b shows the evolution of the gas density for all three models. D16 results in the highest densities in the range of 10$^{11}$ - 10$^{12}$~cm$^{-3}$. These high values stem from a final location in the protoplanetary disc close to the midplane, (z~$\sim$~0.01~au; \citealt{2016MNRAS.462..977D}). As indicated by the evolution of the distance to the protostar, D16 enters a higher density range soon after the onset of the collapse ($\sim$~8.0~$\times$~10$^4$~yr; 0.33~t$_{\rm collapse}^{\rm D16}$). The sooner the trajectories travel inwards, the earlier an increase in density occurs.  After the collapse, the innermost trajectory at 10.8~au shows the highest density of $\sim$~3.4~$\times$~10$^{12}$~cm$^{-3}$, the outermost trajectory at 46.7~au the lowest of $\sim$~4.2~$\times$~10$^{11}$~cm$^{-3}$. This behaviour is not reproduced in the W14-M20 and S20 models, as no protoplanetary disc is considered in these models. The gas density of W14-M20 stays constant during the majority of the collapse, no variation depending on the final distance to the protostellar source is seen. After 3.0~$\times$~10$^5$~yr ($\sim$~0.87~t$_{\rm collapse}^{\rm W14-M20}$), the final values of around $\sim$~10$^9$~cm$^{-3}$ are approached. This indicates, that materials spend most of the collapse phase in the envelope and would only breach into the protoplanetary disc, if it were to be included in the model, at the very end. Similar behaviour is also seen for S20: it takes almost 3.0~$\times$~10$^5$~yr ($\sim$~0.95~t$_{\rm collapse}^{\rm S20}$) for the density to increase to 10$^6$~cm$^{-3}$. The final values of the considered trajectories range between 10$^{8}$-10$^9$~cm$^{-3}$ depending on their distance to the protostar.

\subsubsection{Dust temperature}
The contrasting infall paths of the three physical models are also reflected in the evolution of the dust temperature as depicted in Panel \ref{physics_results}c. The dust temperatures in D16, starting from 10~K, reach their individual maxima around the time when the trajectories breach the innermost 100~au. The innermost trajectory arrives in proximity of the protostar earlier than others, and hence its T$_{\rm dust}$ peaks first. The temperatures of the considered trajectories at the end of the collapse range between $\sim$~20 and 60~K, which is consistent with their positions in the cold, shielded midplane at z~$\sim$~0.01~au (\citealt{2016MNRAS.462..977D}). At the end of the collapse, all have T$_{\rm dust}$ lower than the maxima encountered at disc entry, except for the innermost parcel of material.
\noindent The initial dust temperature in W14-M20 is 10~K for all trajectories at the beginning of the collapse. Given the low density regimes and the large radii, the temperatures remain low for the majority of the collapse. Warmer surroundings ($\sim$~20~-~50~K) are encountered as the emerging protostar heats up the encompassing material after 2.5~$\times$~10$^5$~yr (=~t$_{\rm *birth}^{\rm W14-M20}$). Close to the end of the collapse the innermost trajectory reaches values of $\sim$~270~K, resulting in the highest temperatures of all considered trajectories in this work.
\noindent The dust temperatures calculated with S20 are the lowest of the considered models and share an initial value of 8~K, which remains constant for the majority of the collapse. After the protostar has turned on after $\sim$~3.1~$\times$~10$^5$~yr (=~t$_{\rm *birth}^{\rm S20}$), an increase in temperature is seen. Even though this model remains in the low-temperature (8~K) regime for the longest time, the final temperatures are higher than those in D16 (ranging from 30 to 210~K depending on the proximity to the protostar). Thus, contrasting to D16, T$_{\rm dust}$ at the end of the collapse equals the temperature maxima.
Fig. \ref{physics_results} shows that the distance to the protostar does not correlate with the dust temperature, but rather with the density regime for all models. Although D16's trajectories end up closest to the protostar, the calculated T$_{\rm dust}$ values from S20 and W14-M20 are higher by an order of magnitude (S20 yields a dust temperature of $\sim$ 210~K for the trajectory at 32~au, while D16 has $\sim$~25~K at 30.2~au due to the proximity to the midplane).
The dust temperature correlates with distance to the protostar only as long as the material remains in the envelope with a smooth, radial density profile.

\subsubsection{F$_{\rm UV}$ flux}
The F$_{\rm UV}$ flux is shown in Panel \ref{physics_results}d relative to G$_0$. For S20, the F$_{\rm UV}$ field is assumed to be 10$^{-8}$~G$_0$ at all points in time. The F$_{\rm UV}$ flux in D16 is obtained from calculations with \textsc{Radmc-3d}, as detailed in Section \ref{D16}. 
In this model, the individual F$_{\rm UV}$ maxima of the trajectories correlate with the infall of the material closer to the protostellar source, where the gas density and the dust temperature show a significant increase. All trajectories remain at a constant value of $\sim$~10$^{-6}$~G$_0$ for the first $\sim$~8.0~$\times$~10$^4$~yr (0.33~t$_{\rm collapse}^{\rm D16}$). Afterwards, F$_{\rm UV}$ peaks at individual points in time with the innermost trajectory peaking first at the lowest maximum F$_{\rm UV}$. The two outermost trajectories reach $\sim$~1~G$_0$, due to their entry into the disc being later in time, in comparison to the innermost trajectories, and the protostellar luminosity increasing with time (Fig. 2 of \citealt{2014MNRAS.445..913D}). Afterwards, a decline occurs for all trajectories towards the end of the collapse upon their entry into the shielded protoplanetary disc. Calculations of the F$_{\rm UV}$ flux are not included in model W14-M20, but A$_{\rm V}$ is calculated (Section \ref{W14-M20}). In order to compare the models, the F$_{\rm UV}$ flux shown in Fig. \ref{physics_results} is obtained via the expression F$_{\rm UV}$~=~F$_0$~exp(-$\tau_{\rm UV, eff}$) (\citealt{2011A&A...534A.132V}), where $\tau_{\rm UV, eff}$ is calculated with eq. (\ref{eq_AV}). F$_0$ is set to 1~G$_0$, which assumes an impinging unattenuated UV field of 1~G$_0$ at every position in the envelope. W14-M20 shows the highest F$_{\rm UV}$ flux at the beginning of the collapse with a value of $\sim$~10$^{-6}$~G$_{0}$ for all considered trajectories. A decrease in the flux is seen after $\sim$~2.0~$\times$~10$^5$~yr ($\sim$~0.58~t$_{\rm collapse}^{\rm W14-M20}$). At the end of the collapse, the F$_{\rm UV}$ flux becomes negligible as the parcels are far from the external UV field. In comparison to the initial values, the fluxes at the end of the collapse decrease by about an order of magnitude for D16. Due to the disc entry a spike in the F$_{\rm UV}$ flux occurs for D16, but all trajectories end in the shielded environment of the disc. This decrease from initial to final also occurs for W14-M20; however, for a different reason that is the distancing from the externally impinging UV radiation.

\subsection{Tracing the physical components in the models}\label{sec_tracing_components}
As discussed in Section \ref{Results_physics_text}, trajectories pass through various physical components of a low-mass star-forming region during their infall towards the central source. Depending on the model, this can include components such as the inner and outer envelope, protoplanetary disc, and outflow cavities. Here, all three models are described in the context of the generic 2D D16 model.
\newline 
\noindent \subsubsection{Tracing the physical components in the models}
\begin{figure}
    \centering
    \includegraphics[width=1.0\columnwidth]{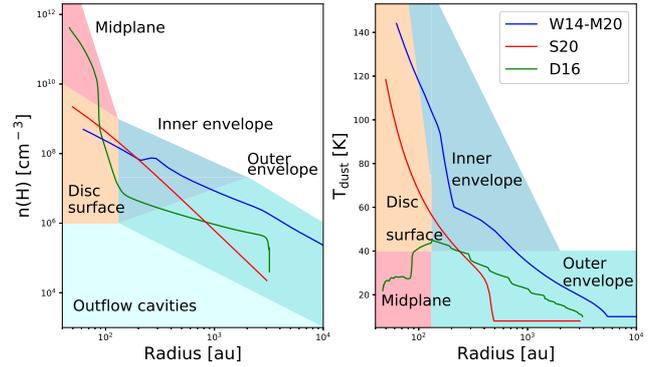}
    \caption{Gas density relative to n$_{\rm H}$ in cm$^{-3}$ (left panel) and dust temperature regimes in Kelvin along the infall paths of the discussed trajectories as a function of distance to the protostar. The W14-M20 model is plotted in blue, S20 in red, and D16 in green. The shaded regions correspond to regions of a star-forming system.}
    \label{TdnH2_vs_R}
\end{figure}

Fig. \ref{TdnH2_vs_R} addresses the components of low-mass star formation (Section \ref{sec_tracing_components}) encountered by three trajectories of the models studied in this work. The final positions of the trajectories considered are 46.7~au (D16), 49.7~au (S20) and 62.4~au (W14-M20) and are also later used to study the chemical evolution in detail. The radial position of the infalling trajectories is depicted as a function of gas density (left panel) and dust temperature (right panel). The depicted parameter ranges of the shaded regions are taken from the D16 model as detailed in Fig. 5 and Section 3.1.2 in \cite{2014MNRAS.445..913D} and resemble the description of the 2D model when the trajectory of 46.7~au enters the forming protoplanetary disc. The size of the protoplanetary disc is set to 130~au, but it extends further as the system evolves. The outer envelope is characterised by n$_{\rm H}$~=~10$^6$~-~10$^7$~cm$^{-3}$ and T$_{\rm dust}$~<~40~K with a radial distance beyond 130~au. The same radial distance, but densities between 10$^6$ and 10$^9$~cm$^{-3}$ and T$_{\rm dust}$~>~40~K are attributed to the inner envelope. The protoplanetary disc regime accounts for the disc surface and the midplane: the disc surface covers regions with n$_{\rm H}$ from 10$^6$~-~10$^{\rm 10}$~cm$^{-3}$ and T$_{\rm dust}$~$>$~40~K, the midplane shows densities of at least 10$^9$~-~10$^{\rm 10}$~cm$^{-3}$ depending on the radial distance and T$_{\rm dust}$~$<$~40~K. The outflow cavities can extend over the full considered radial distance of the model and show n$_{\rm H}$ between 10$^4$ and 10$^6$~cm$^{-3}$ and T$_{\rm dust}$~$\sim$~90~-~300~K.
\newline
\noindent As shown by the different traced density and dust temperature regimes in Fig. \ref{TdnH2_vs_R}, it becomes evident, that the three models are constructed for different regions of the forming system. 
All models start their infall paths in the outer envelope, their infalls are accompanied by increasing n$_{\rm H}$, as they pass the inner envelope and approach their final positions in the disc regime. The 1D nature of W14-M20 and S20 combined with being tailored towards IRAS 16293 results in very similar gas densities at their final positions ($\sim$~10$^9$~cm$^{-3}$). Accounting for the 2D structure of the system, D16 traces the trajectory as it approaches the midplane at z~$\sim$~0.01~au, which results in a gas density at the final position on the order of 10$^{\rm 11}$~cm$^{-3}$.
Due to the deviating t$_{\rm *birth}$, the infalling material covers distinct distances unperturbed by the influence of the protostar. As soon as the models evolve past t$_{\rm *birth}$, T$_{\rm dust}$ starts to increase. D16 is already influenced at the beginning of its infall path. W14-M20 travels $\sim$~5~000~au before the protostar emerges. Afterwards, T$_{\rm dust}$ increases up to $\sim$~140~K at its final position. The steady increase is given due to the parameterisation of T$_{\rm dust}$ in terms of radius (Section \ref{W14-M20}). S20 lies only at a distance of 500~au at t$_{\rm *birth}$, then T$_{\rm dust}$ increases to $\sim$~120~K. The final position in the shielded midplane for D16 is accountable for the T$_{\rm dust}$ of $\sim$~20~K at the end of the calculations, its temperature maximum of $\sim$~45~K is obtained prior to the disc entry. 2D models allow a more comprehensive exploration of the different physical components of a star-forming system and cover a more complete range of encountered values of physical parameters.

\subsection{Initial atomic abundances}
\begin{figure}
    \centering
    \includegraphics[width=1.0\columnwidth]{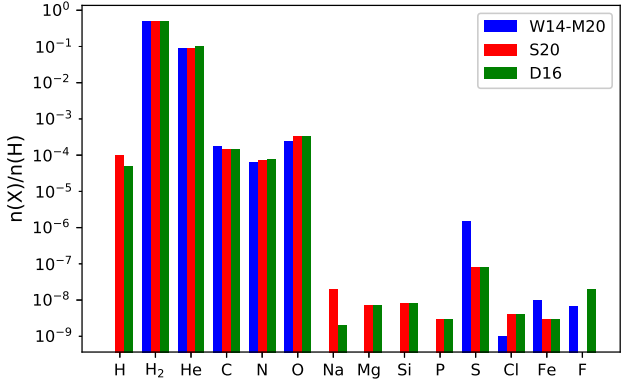}
    \caption{Initial atomic abundances (Table \ref{initial_abundances}) at the start of the precollapse phase relative to n$_{\rm H}$. Blue bars correspond to the values of the W14-M20 model, red bars depict the S20 model, and green bars refer to the D16 model. All species first start in the gas. In the case of W14-M20, C, S, Fe, and Cl are initially purely ionic.}
    \label{initialatomic}
\end{figure}

The initial atomic abundances at the start of the prestellar core phase are plotted in Fig. \ref{initialatomic} (values given in Table \ref{initial_abundances}). In the case of W14-M20, C, S, Fe, and Cl are initially purely ionic. For this work, the most important elements are atomic H and molecular H$_{\rm 2}$, and C, N, O, and S. The values for C, N, and O concur for all three models: their values differ by less than a factor of 10. However, the initial abundance of sulfur differs by two orders of magnitude (S$^+$/n$_{\rm H}$~$\sim$~10$^{-6}$ for W14-M20, S/n$_{\rm H}$~$\sim$~10$^{-8}$ for D16 and S20). Another difference between the models is the value of atomic hydrogen. While D16 contains 10$^{-5}$, S20 begins with  10$^{-4}$. In W14-M20, all H is initially in its molecular form. The overall number of hydrogen nuclei (n$_{\rm H,nuclei}$) is given by n$_{\rm H,atomic}$+2n$_{\rm H_2}$. Availability of atomic hydrogen is important for hydrogenation reactions on the surfaces of grains. All models also include elements that are not used to produce the molecules central to this work. Fluorine is included only in W14-M20 and D16, Fe and Cl are present in all three models. Furthermore, S20 and D16 include Na, Mg, Si, and P, which all lie in the range of 10$^{-8}$ - 10$^{-9}$ relative to n$_{\rm H}$. In the case of D16, all values are taken from the UMIST database, which account for typical abundances in dark cloud cores (see \citealt{2013A&A...550A..36M} for details). The values used by the S20 model with the exception of H and H$_2$ have already been published in \cite{2013ApJ...765...60G}.

\subsection{Precollapse molecular abundances}\label{sectionprecollapse}
\begin{figure}
    \centering
    \includegraphics[width=1.0\columnwidth]{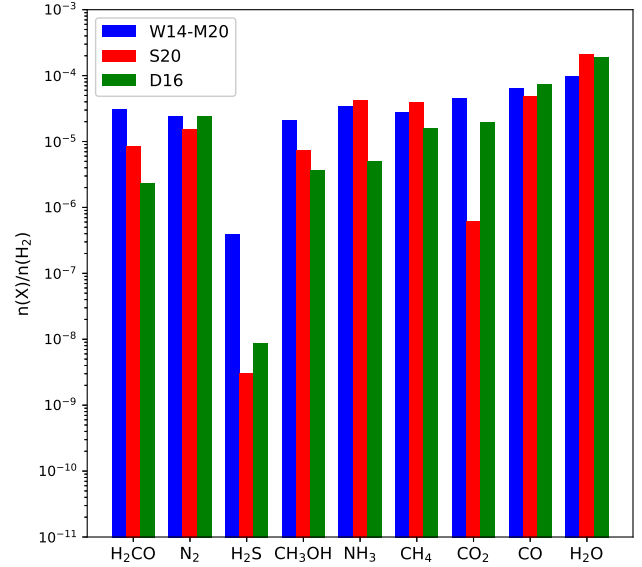}
    \caption{Precollapse molecular abundances (Table \ref{initial_molecular}) at the start of the collapse phase relative to n$_{\rm H_{2}}$. Blue bars describe the values of the W14-M20 model, red bars show the S20 model, and green bars correspond to the D16 model. This plot depicts the sum of gases and ices (individual phases are shown in Fig. \ref{initialmol}).}
    \label{initialmol_sum}
\end{figure}

The precollapse molecular abundances are plotted in Fig. \ref{initialmol_sum} and Fig. \ref{initialmol} (values listed in Table \ref{initial_molecular}).  To allow a proper comparison of the three-phase versus the two-phase chemical models, the abundances in the bulk and on the surface are summed to obtain the total abundances in the ice. Furthermore, to highlight chemical evolution rather than the partition between phases, the gas and ice abundances have been summed. Due to the cold dust temperatures ($\sim$~10~K) of the precollapse phase, the molecules in Fig. \ref{initialmol_sum} are predominantly found as solids (Fig. \ref{initialmol}). Good agreement between models is obtained for the hypervolatile (\cite{2006A&A...449.1297B}) species N$_2$ and CO (within a factor of $\sim$~1.4~\textendash~1.6), and also for H$_2$O and CH$_4$ (factor of 1.1~\textendash~2.6). In the case of CO$_2$, two models (W14-M20 and D16) produce almost identical numbers (within a factor of 2), but S20 lies two orders of magnitude lower. This is a bit puzzling, as the species that play an important role in the formation of CO$_2$ (specifically, CO and H$_2$O; \citealt{2011ApJ...735..121N}) show comparable numbers in all the models. CO$_2$ forms on grain surfaces via the reaction of CO and OH, with OH originating from the photodissociation of H$_2$O, when dust temperatures are sufficiently low to keep CO in the solid phase (\citealt{2016MNRAS.462..977D}). The underproduction of CO$_2$ in S20 may be explained by the small overproduction of H$_2$O and CH$_4$ in that model, which decreases the availability of C and O to form CO$_2$. Another possibility is the lack of UV in model S20, which limits the amount of OH stemming from the photodissociation of H$_2$O. Thus, less OH is available to form CO$_2$ via the reaction of CO and OH. For NH$_3$, the abundance in D16 is an order of magnitude lower than in W14-M20 and S20. In the case of H$_2$CO, and CH$_3$OH, the values lie within the same order of magnitude, but with D16 having the lowest abundances. The D16 model has the shortest precollapse timescale (Table \ref{initial_parameters_physics}), thereby reducing the amount of time for species such as NH$_3$, H$_2$CO, and CH$_3$OH to form via hydrogenation reactions (\citealt{Ioppolo2011}, \citealt{2015MNRAS.446..439F}). Consequently, t$_{\rm precollapse}$ is a critical parameter for molecules formed in prestellar cores. However, if initial atomic abundances vary by more than an order of magnitude, then this will also affect the molecules formed. This is the reason for the big differences in the abundance of H$_2$S: $\sim$~10$^{-7}$ (W14-M20) over 10$^{-8}$ (D16) to 10$^{-9}$ (S20) relative to n$_{\rm H_{2}}$.

\subsection{Post-collapse molecular abundances}\label{postcollapseabundances}
In order to compare the chemical evolution during the collapse as computed by the three physicochemical models, one trajectory per model is selected. The final positions considered are 46.7~au (D16), 49.7~au (S20), and 62.4~au (W14-M20). Fig. \ref{species_sum} depicts the sum of the gas and ice abundances during the collapse along these three trajectories of the three physicochemical models analyzed in this work. Fig. \ref{H2CO}-\ref{H2O} contain the analogous figures for the two phases separately on a molecule by molecule basis.

\subsubsection{Similarities}\label{similarities}
W14-M20 and S20 predominantly trace the chemistry in the envelope at large scales from the forming protostar. Fig. \ref{species_sum} shows that the prestellar abundances remain unaltered for the majority of the collapse (for t~$<$~0.95~t$_{\rm collapse}^{\rm S20}$ and t~$<$~0.87~t$_{\rm collapse}^{\rm W14-M20}$, respectively) with no chemical evolution of the material taking place. In comparison to D16, this is only reproduced for N$_2$ and H$_2$O (although, even H$_2$O shows a relatively small response to the disc entry, Section \ref{discentry}). 

\subsubsection{Differences arising from the precollapse molecular abundances}\label{differences}
At the onset of collapse the abundances of H$_2$S and CO$_2$ differ by several orders of magnitude between the models (Section \ref{sectionprecollapse}). These differences are preserved during the collapse and at the end (Fig. \ref{species_sum} and \ref{env_vs_disc}). Both of these molecules do also undergo some additional chemical processing during the collapse, most noteworthy in D16. This is also seen for the case of H$_2$CO, but to a lesser extent.
\newline 
\noindent The abundance of H$_2$S is lowered at some point during the collapse in all models. In D16, it decreases by three orders of magnitude upon disc entry (Section \ref{discentry}) due to photodissociation, but it is reformed quickly once inside the disc up to its envelope abundance. In S20, H$_2$S begins to undergo destruction at times greater than t$_{\rm *birth}$ due to photodissociation.
\newline
\noindent For all models, the evolution of the CO$_2$ abundances display the same trend: it increases during the collapse (by a factor of $\sim$~5 for W14-M20, $\sim$~75 for S20, and $\sim$~400 for D16). 
\newline
\noindent H$_2$CO increases its abundance towards the end of the collapse for W14-M20 and S20, the difference between the models is an attribute of the abundance at the start of the collapse. While it undergoes additional processing in D16 (Section \ref{discentry}), its maximum throughout the collapse never exceeds its precollapse molecular abundance, which is about an order of magnitude lower than for the other two models as discussed in Section \ref{sectionprecollapse}.

\subsubsection{Strong influence by the disc entry}\label{discentry}
\begin{figure*}
    \centering
    \includegraphics[width=1.0\textwidth]{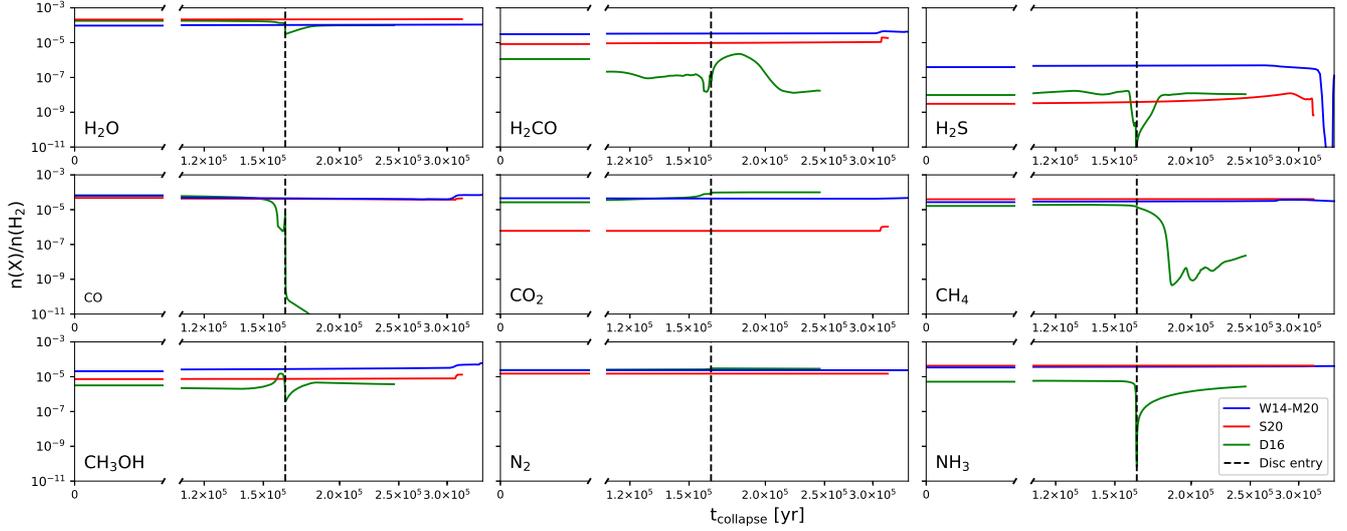}
    \caption{This figure depicts the sum of the abundances in the gas and ice relative to n$_{\rm H_{\rm 2}}$ for all molecular species throughout the collapse. Blue represents W14-M20 (final position at 62.4 au), red S20 (final position at 49.7 au),  and green D16 (final position at 46.7 au). The dashed black line marks the disc entry of the D16 model.}
    \label{species_sum}
\end{figure*}

Given the 1D nature of W14-M20 and S20 representative of the envelope, the impact of the disc entry can only be assessed for D16. The trajectory enters the protoplanetary disc at $\sim$~1.65~$\times$~10$^5$~yr ($\sim$~0.67~t$_{\rm collapse}^{\rm D16}$).
The disc is characterized by cold ($<$~50~K) dust temperatures, high densities ($>$~10$^9$~cm$^{-3}$), and small radii ($<$~100~au). This results in a rapid change in physical conditions experienced by the infalling material from the envelope.
Fig. \ref{species_sum} shows that the majority of the species are heavily impacted by the disc entry in D16: the changes seen in the abundances of CO, H$_2$CO, and CH$_3$OH are the result of hydrogenation reaction rates being strongly affected in the range of dust temperatures and densities experienced at the disc entry (\citealt{2009A&A...505..629F}). Photodissociation due to the strong F$_{\rm UV}$ in the inner envelope causes a large decrease in the abundances of NH$_3$ and H$_2$S prior to the disc entry. This is also the reason for the small change in the abundance of H$_2$O (Section \ref{similarities}). This, in turn, leads to the production of CO$_2$, making CO$_2$ more abundant in D16 than in the other two physicochemical models (Section \ref{differences}). The photostable N$_2$ (\citealt{2013A&A...555A..14L}) is the only molecule that is not affected by the disc entry (Section \ref{similarities}).

\subsubsection{Young vs. mature envelope}\label{sec_discvsenv}
\begin{figure}
    \centering
    \includegraphics[width=1.0\columnwidth]{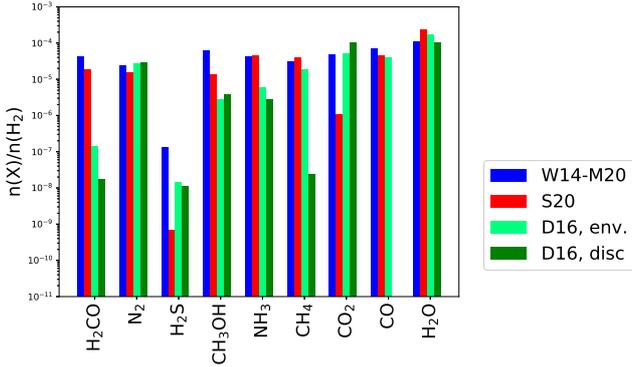}
    \caption{Post-collapse molecular abundances relative to n$_{\rm H_{2}}$. Blue bars describe the values of the W14-M20 model (final position at 62.4 au), red bars show the S20 model (final position at 49.7 au), light green bars correspond to the abundances in the warm envelope upon disc entry in the D16 model (at t~=~1.5~$\times$~10$^5$~yr =~0.61~t$_{\rm collapse}^{\rm D16}$), dark green bars illustrate the abundances in the disc at the end of the collapse in the D16 model (final position at 46.7 au).}
    \label{env_vs_disc}
\end{figure}

Fig. \ref{env_vs_disc} shows the abundances of nine selected molecules at the respective t$_{\rm collapse}$ of the three physicochemical models, W14-M20, S20, and D16. As discussed in Section \ref{discentry}, the disc entry significantly alters the abundances of almost all molecules. Consequently, in Fig. \ref{env_vs_disc}, a comparison is also made to the envelope abundances of D16 that are obtained before the trajectory enters the disc. The envelope abundances of D16 are taken at an age of 1.5~$\times$~10$^5$~yr ($\sim$~0.61~t$_{\rm collapse}^{\rm D16}$), which implies that it is a younger envelope in comparison to the mature envelopes of W14-M20 and S20 at an age of t$_{\rm collapse}$. The agreement between models for N$_2$ and H$_2$O remains also for the envelope (Sections \ref{similarities} and \ref{discentry}). CO$_2$ and H$_2$S differences persist as a result of their initial precollapse abundances (Section \ref{differences}). Some of the hydrogenation-dominated species (CO, H$_2$CO, NH$_3$; Section \ref{discentry}) now show a closer agreement in the envelope within a factor of 2 for CO and 10$^2$ for H$_2$CO (compared to a factor of 10$^3$ for the post-collapse abundance). NH$_3$ now shows an agreement within a factor of 7 as the envelope abundance is its abundance prior to photodissociation at disc entry (Section \ref{discentry}). The CH$_3$OH abundance remains in close agreement between the models (with the envelope abundance being a mere factor of 1.4 lower than the abundance at t$_{\rm collapse}$). Finally, CH$_4$ also drastically improves in terms of agreement as the envelope abundance is almost three orders in magnitude higher than in the disc in D16. The younger age of the D16 envelope is the reason for the envelope abundances being systematically lower for hydrogenation-dominated species (CO, H$_2$CO, CH$_3$OH, NH$_3$). CO$_2$ is the only species more abundant in the envelope of D16 than in W14-M20 and most-notably in S20 due to significant F$_{\rm UV}$ flux in the inner envelope at lukewarm temperatures, which is not a regime covered in the other two models (Sections \ref{similarities} and \ref{differences}). Therefore, the envelope age is important for interpreting chemical outputs of physicochemical models, as well as the rate at which the physical parameters are evolved.

\subsection{Molecular abundances with the same chemical model}

Dismantling all three chemical networks that all include hundreds of species and thousands of reactions is not feasible. Instead, the molecular abundances are re-calculated with the two-phase chemical code of D16 for the precollapse physical parameters of W14-M20 and S20. Although the precollapse abundances are now calculated with the same chemical code, differences in the molecular budget at the onset of the collapse remain distinct due to the differences in the physical parameters. These are compared in Section \ref{sec_precoll_redone}. For the results presented in Section \ref{sec_postcoll_redone}, the adopted prestellar physical parameters and abundances of the D16 model are taken as identical initial conditions for the W14-M20 and S20 models as well (Table \ref{initialmol}). This approach allows tracking the difference in the chemical evolution solely based on the input provided by the physical model of the collapse phase.

\subsubsection{Precollapse molecular abundances}\label{sec_precoll_redone}

\begin{figure}
    \centering
    \includegraphics[width=1.0\columnwidth]{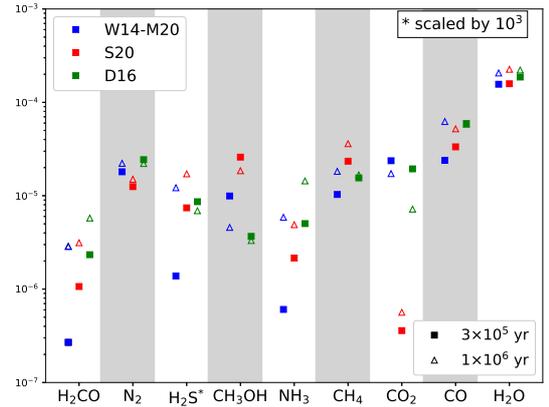}
    \caption{Precollapse molecular abundances at the start of the collapse phase relative to n$_{\rm H_2}$ for the considered trajectories calculated with the chemical model of D16 (gases and ices have been summed). Blue markers describe model W14-M20, red markers present S20, and green markers depict D16. The fiducial D16 model for t$_{\rm precollapse}^{\rm D16}$~=~3.0~$\times$~10$^5$~yr is indicated by the filled green squares. Filled blue and red squares are used to describe W14-M20 and S20 for the same precollapse duration. Triangle markers describe a precollapse duration of 1.0~$\times$~10$^6$~yr. Note that the abundances of H$_2$S are scaled by a factor of 10$^3$ for all models.}
    \label{fig_precollapse_recalc}
\end{figure}

The recalculations of the chemical evolution during the precollapse stage are depicted in Fig. \ref{fig_precollapse_recalc}. Calculations are performed for precollapse durations of 3.0~$\times$~10$^5$~yr and 1.0~$\times$~10$^6$~yr for all three models. While the individual physical precollapse parameters are implemented (Table \ref{initial_parameters_physics}), only the initial atomic abundances of the D16 model (Table \ref{initial_abundances}) are considered. 
\newline
Running the same chemical model for the same precollapse duration shows that the marginally different precollapse parameters still affect the outcome of the molecular abundances. Neither for a precollapse duration of 3.0~$\times$~10$^5$~yr (indicated by filled squares in Fig. \ref{fig_precollapse_recalc}) nor for a precollapse duration of 1.0~$\times$~10$^6$~yr (indicated by hollow triangles in Fig. \ref{fig_precollapse_recalc}), is a perfect match obtained for any single molecule. This shows that prestellar density, dust temperature, and extinction values are critical to the highest precision. In turn, the precollapse duration is the most critical physical parameter. While the spread between the models for the same duration is significant, the abundance differences for different durations can be orders of magnitude apart. With the exception of CH$_3$OH, the longer precollapse duration leads to a higher abundance of the species formed via hydrogenation reactions on the grain surfaces (CH$_4$, NH$_3$, H$_2$O, and H$_2$CO). This hints that COM formation already occurs efficiently during the precollapse phase. Production of H$_2$CO, CH$_3$OH, and COMs via efficient hydrogenation explains the reduction of CO$_2$ for t$_{\rm precollapse}$~=~10$^6$~yr for models D16 and W14-M20. The longer precollapse duration also increases the CO abundances for S20 and W14-M20, in the case of D16 the value does not change.

\subsubsection{Post-collapse molecular abundances}\label{sec_postcoll_redone}

\begin{figure*}
    \centering
    \includegraphics[width=1.0\textwidth]{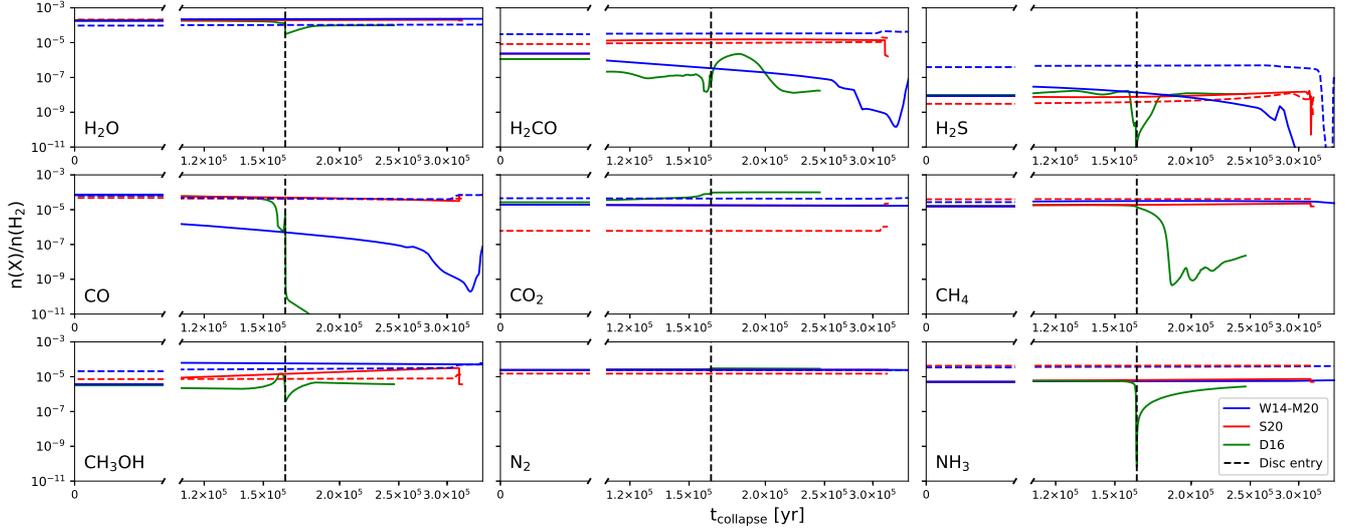}
    \caption{This figure depicts the sum of the abundances in the gas and ice relative to n$_{\rm H_{\rm 2}}$ for all molecular species throughout the collapse. Blue represents W14-M20 (final position at 62.4 au), red S20 (final position at 49.7 au), and green D16 (final position at 46.7 au). The dashed black line marks the disc entry of the D16 model. The solid lines represent results obtained with the two-phase chemical network. The dashed lines corresponds to results from \textsc{Nautilus} (blue) and \textsc{Magickal} (red) and are depicted as a reference.}
    \label{species_sum_redone}
\end{figure*}

For the calculation of the chemical evolution during the collapse, the precollapse molecular abundances of D16 (Table \ref{initial_molecular}) are used as a starting point to remove the influence of the different abundances at t$_{\rm  collapse}$~=~0. The chemical evolution during the collapse as calculated with the chemical code of D16 for trajectories of all the physical models considered in this work is depicted in Fig. \ref{species_sum_redone} with solid lines. As a reference for the reader, the dashed lines correspond to the original results obtained with \textsc{Nautilus} and \textsc{Magickal} for the W14-M20 and S20 trajectories, respectively. Recalculating the chemical evolution allows to split the molecular abundances into two distinct groups.
\newline
\noindent \textit{Variations mainly due to the now-identical adopted precollapse abundances:} Fig. \ref{species_sum_redone} shows that the evolution of N$_2$, H$_2$O, CH$_4$, NH$_3$, H$_2$S, and CO$_2$ are hardly impacted by the switch of the chemical network (less than a factor of 10). This is also true for CO in S20. The observed shift in abundances between the old and new calculations stems only from the now different molecular abundances at the beginning of the collapse. This shows that neither the newly adopted binding energy of $0.3$~$\times$~E$_{\rm des}$ (in contrast to $0.35$~$\times$~E$_{\rm des}$ for S20 and $0.4$~$\times$~E$_{\rm des}$ for W14-M20) nor the different number of reactants and reactions influence these molecules at cold conditions before t$_{\rm *birth}$ and after, as they enter the hot corino and encounter warmer regions with higher gas densities. For these species, the abundance during the collapse can be computed accurately with two- and three-phase chemical models once the initial precollapse abundance is set. 
\newline 
\noindent \textit{Variations due to the choice of the chemical network:}
Major differences between the results obtained with the two-phase and three-phase chemical models arise in species that are critical in the CO hydrogenation sequence towards H$_2$CO and CH$_3$OH. In comparison to the three-phase calculations, the following results are obtained with the two-phase computation: the CO abundance is lower for S20 and W14-M20; the H$_2$CO abundance is higher for S20 and lower for W14-M20; the CH$_3$OH abundance is higher for W14-M20 and S20. For the case of S20 (\textsc{Magickal}), the differences are typically less than one order of magnitude. However, for the case of W14-M20 (\textsc{Nautilus}), the differences for CO and H$_2$CO can be as large as 3 orders of magnitude. The relatively small differences between the two-phase model and S20 (\textsc{Magickal}) suggest that two-phase models do not necessarily overproduce reactivity in the solid phase. This is supported by recent findings of \cite{2016A&A...591A...9S}, that show that bulk ice models are better suited to describe the chemistry in starless cores. The source of the large discrepancies between models may stem from adopted activation barriers in individual reactions. These are then carried forward throughout the entire duration of the collapse. Differences seen in Fig. \ref{env_vs_disc} between the postcollapse abundances of the three models are reduced, as expected, upon computation with an identical chemical network (Fig. \ref{env_vs_disc_newchem}).

\section{Discussion}\label{Discussion}

\subsection{Timescales}\label{sec_timescales}

For the three studied models, the precollapse phase lasts either 3~$\times$~10$^5$~yr (D16) or 10$^6$~yr (W14-M20 and S20; Table \ref{initial_parameters_physics}). Dynamical evolution is proposed for cloud cores in the turbulent paradigm with a lifetime of 1~-~10 free-fall times (t$_{\rm ff}$):
\begin{equation}
    t_{\rm ff} = \sqrt{\frac{3\pi}{32G\rho}} =
    \sqrt{\frac{3\pi}{32Gm_{\rm H} \mu _{\rm p} \langle n \rangle}},
\end{equation}
where G is the gravitational constant, $m_{\rm H}$ the mass of the hydrogen atom, $\mu _{\rm p}$ the mean molecular weight per particle, and $\langle n\rangle$ the average gas density (\citealt{2009sfa..book..254A}). For the models considered in this work, t$_{\rm ff}$ spans between $\sim$~1.7~$\times$~10$^5$~yr (t$_{\rm ff}^{\rm D16}$) and 3.17~$\times$~10$^5$~yr (t$_{\rm ff}^{\rm S20}$). Thus, the precollapse timescales of 3.0~$\times$~10$^5$ ($\sim$~1.8~t$_{\rm ff}^{\rm D16}$) and 1.0~$\times$~10$^6$~yr ($\sim$~3.2~t$_{\rm ff}^{\rm S20}$) lie well within the theoretical 1~-~10~t$_{\rm ff}$ range.
\newline 
\noindent Observations of dense cores indicate that their lifetimes depend on mass. Studies of isolated cores have shown, that their lifetimes as starless cores decrease as the density increases (\citealt{2000MNRAS.311...63J}). Observations of \cite{1999ApJ...526..788L} suggest that the typical lifetime of a starless core with an average density of $\sim$~10$^4$~cm$^{-3}$ should be on the order of 1.0 - 1.5~$\times$~10$^6$~yr. This is longer than the timescale used in D16, but the initial density in this model is also higher than 10$^4$~cm$^{-3}$. Observations of prestellar cores in Perseus, Serpens, and Ophiucus show that the ratio of starless to protostellar cores in each cloud is close to unity, which suggests, that the core lifetime before the onset of collapse should be similar to the lifetime of the embedded protostar. This leads to an average prestellar lifetime of $\sim$ 4.5~$\times$~10$^5$~yr with an uncertainty of a factor of 2, which is also in good agreement with the models presented here  if one assumes that all starless cores eventually will turn prestellar (\citealt{2008ApJ...684.1240E}).
\newline 
\noindent Even though the different t$_{\rm precollapse}$ of the models all lie within an acceptable range of the observations and theory, the difference of 7.0~$\times$~10$^5$~yr still impacts the precollapse molecular budget (Fig. \ref{initialmol_sum} and \ref{initialmol}; Section \ref{sectionprecollapse}). Another relevant timescale is t$_{\rm collapse}$, which is set to 2.5~$\times$~10$^5$ yr (D16), 3.4~$\times$~10$^5$~yr (W14-M20), and 3.2~$\times$~10$^5$~yr (S20). Observations find that the collapse, if turbulence is the primary driver, should last 1~-~2~t$_{\rm ff}$ (\citealt{2008ApJ...684.1240E}). This is in agreement with W14-M20 and S20, as their respective t$_{\rm collapse}$ equal 1~t$_{\rm eff}$ (S20) or fall in the theoretically determined range of 1~-~2~t$_{\rm ff}$ ($\sim$~1.44~t$_{\rm ff}^{\rm W14-M20}$). The D16 model (t$_{\rm collapse}^{\rm D16}$~$\sim$~1.46~t$_{\rm ff}^{\rm D16}$) also fits the observational evidence, even though this model does not assume free-fall collapse (Section \ref{D16}). However, quantifying the contraction motions and thus the dynamical evolution can only be achieved by carrying out a detailed radiative transfer study. This allows to compare computed line profiles with observational ones (e.g., \cite{2008ApJ...683..238K} for prestellar cores). Furthermore it is to note, that t$_{\rm *birth}$ is close to t$_{\rm ff}$ for W14-M20 ($\sim$~0.95~t$_{\rm ff}^{\rm W14-M20}$) and S20 ($\sim$~0.97~t$_{\rm ff}^{\rm S20}$), but this is not the case for D16 ($\sim$~0.12~t$_{\rm ff}^{\rm D16}$). The moment of stellar birth determines the onset of active chemistry during the process of star formation (Section \ref{sec_discvsenv}) stimulated by higher dust temperatures and enhanced F$_{\rm UV}$ fluxes.

\subsection{Physical components of low-mass star-forming regions}\label{section_components}
\noindent It is not easy to disentangle the envelope and the disc of young embedded protostars in observations. Traditionally, protostars are classified by their spectral energy distribution (SED; \citealt{1987IAUS..115....1L}). The earliest Class 0 stage is deeply embedded, which makes observations at near-infrared (NIR) wavelengths difficult. Class I objects are still embedded in an envelope that feeds the protoplanetary disc and accretes onto the protostar, but these objects are more easily accessible in the NIR. In Class II objects, the envelope has dissipated, but the protoplanetary disc still accretes onto the protostar. Observations suggest, that this phase sets in $\sim$~5~$\times$~10$^5$~yr after the collapse (\citealt{2014prpl.conf..195D}). The presence of the protostellar envelope at earlier times does not yet allow a direct comparison of discs associated with Class I to Class II objects for a big sample. Estimates from observations attribute radii as large as $>$~250~au to them (\citealt{2014prpl.conf..195D}, \citealt{2014prpl.conf..173L}). This is consistent with D16 (R$_{\rm disc}$~$\sim$~300~au at t$_{\rm collapse}$). In Class 0 objects, 90~\% of the NIR emission stems from the envelope. Consequently, if a disc-like structure exists at this stage, its characterisation is very complicated. In recent years high spatial resolution observations performed with ALMA have shown that COMs do already exist in very young discs (e.g., \citealt{2019ECS.....3.2110C,2019NatAs...3..314L}). The increasing availability of data for the inner parts of protostellar systems will allow us to understand the detailed structures of these sources at different scales. Lifetime estimates of the Class 0 objects from observations suggest that this phase lasts between 1.5-1.6~$\times$~10$^5$~yr (\citealt{2014prpl.conf..195D}). As the envelope is present until the end of the calculations in all models, while the age is greater than 1.6~$\times$~10$^5$~yr, the modelled systems correspond to Class I objects at t$_{\rm collapse}$ following the observational classification. Theoretical work of \cite{2006ApJS..167..256R} classified evolutionary stages according to the accretion rates from the envelope and the disc onto the protostar. This formalism is adapted for D16 by \cite{2013A&A...555A..45H}: Stage 0 is characterized by M$_{\rm env}$~$>>$~M$_{*}$ (t/t$_{\rm collapse}$~$\leq$~0.5), Stage I by M$_{\rm env}$~$<$~M$_{*}$, but M$_{\rm env}$~$>$~M$_{\rm disc}$ (t/t$_{\rm collapse}$~$>$~0.5), and Stage II by M$_{\rm env}$~$<$~M$_{\rm disc}$, which puts all models into the Stage I category at the end of the calculations.

\subsection{Influence of the physical conditions on the chemical model}
Given the similarities of models S20 and W14-M20 (Fig. \ref{physics_results}), the differences when using the same chemical model should not be as prominent as in Fig. \ref{species_sum_redone}, if the role of the physical conditions would not be critical. While the difference in values for, e.g., gas densities or dust temperatures might seem small and be well within the error limits, this will impact the results obtained with chemical networks on timescales applicable to star-forming systems. Cold temperatures at early stages allow gas-phase species to freeze out onto the dust grains, which is further enhanced by a high gas density. The decreasing abundances of gaseous species (e.g., CO) change the composition of the gas and thus the chemistry, as they become available for chemical processes on the grain surfaces instead. Theoretical and experimental work by \cite{2009A&A...508..275C} and \cite{2009A&A...505..629F} on the hydrogenation process from CO to CH$_3$OH show that the obtained abundances and formation efficiencies of methanol and its intermediate products vary significantly within the considered narrow temperature range of 12 \textendash ~20~K and long interstellar timescales. As these variations are already significant in a system with a small number of species and possible reactions, their significance should not be underestimated in a large chemical network. 
\newline 
\noindent Another parameter that can influence the chemical evolution and thus, should not be neglected is the CR ionisation rate. Pioneering modelling work has been performed by \cite{2016A&A...590A...8P}, where the first comprehensive theoretical model of CR production and acceleration along protostellar jets and outflows is presented. Most molecular clouds are thought to have an ionisation rate that varies spatially and can locally be enhanced by a few orders of magnitude as suggested by observations of, e.g., \cite{2014ApJ...790L...1C}, \cite{2014A&A...565A..64P}, \cite{2017A&A...605A..57F}, and \cite{2018ApJ...859..136F}. This has been reproduced by semi-empirical modelling work (\citealt{2019ApJ...878..105G}). However, most models assume a standard CR ionisation rate, which is the case for S20 and W14-M20. Moreover, CRs can penetrate the inner, dense parts, where species freeze out onto the dust grains that are shielded from interstellar F$_{\rm UV}$ and protostellar FUV and X-ray photons. While these species cannot thermally desorb due to the cold temperatures, CRs can hit dust grains and return these species into the gas phase via non-thermal desorption processes and localised thermal events (spot heating; \citealt{1985A&A...144..147L}). CRs can also interact with H$_2$ and produce FUV photons, which can subsequently lead to photodesorption from the grain surfaces (\citealt{2009A&A...496..281O}, \citealt{2009ApJ...693.1209O}) and photodissociation in the gas and directly in the ice. Furthermore, it has been shown that CR interactions can increase the temperature inside the cloud and consequently influence the gas-phase chemistry (\citealt{2017ApJ...839...90B}). Non-thermal chemistry induced by CRs has also been shown to produce COMs under cold  core conditions (\citealt{2018ApJ...861...20S}). In addition CRs are believed to control the fractional abundance of hydrogen atoms in molecular clouds, which are then again important for, e.g., hydrogenation reactions on grain surfaces. The atomic H abundance can be explained if H$_2$ is dissociated by CRs. As shown by \cite{2018A&A...619A.144P}, secondary electrons stemming from the primary ionisation via CRs are the only source of atomic hydrogen at column densities representative of these environments. Either these secondary processes, a variable CR ionisation rate as a result of CR attenuation, or spatially differing CR ionisation rates will therefore impact the results of a chemical model. This is especially true if one considers that some models, as it is in the case of W14-M20, only account for molecular hydrogen, H$_2$, at the beginning of the calculations. 
\newline 
\noindent Lastly, the considered timescales have shown to play the dominant role in the outcome of chemical models. While the duration of the precollapse phase and duration prior to the birth of the protostar have already been discussed in Section \ref{sec_timescales}, one further caveat is present in all three discussed models and common to astrochemical models in general: the chemical evolution is traced throughout the collapse phase and then a few $\times$~10$^5$~yr at most after the birth of the protostar. These results are then used to infer the history of observed systems. However, it is unclear what the appropriate collapse time for each system is, it may be shorter or longer than the model time. Furthermore, it is hard to quantify the age of these systems accurately and consequently it is difficult to constrain how long the abundances after t$_{\rm *birth}$ should be traced with the models.

\subsection{On the need for generic models} 
Physicochemical models can be a powerful predictive tool. In preparation for observations, they can deliver valid estimates of the expected chemistry if the physical conditions of the region of interest are well constrained. On the other hand, they make use of molecular inventories obtained from observations to predict the evolutionary history of the system. The results presented in the previous sections show, that this can be achieved with different modelling approaches. Both 1D and 2D approaches to the physical modelling part, and two-phase and three-phase approaches to the chemical part derive similar results if they probe comparable physical conditions for the majority of the species considered in this work. 
However, 1D models do oversimplify the physical environment, even if some of the physical input, for example, the gas density in the case of W14-M20, or the envelope model for S20, are inferred from observations of a specific source. For observations with a low spatial resolution, where the majority of the probed material corresponds to the envelope; and the hot corino region is not resolved sufficiently; an adequate fit can still be obtained. However, in recent years facilities such as ALMA offer the possibility to constrain the structure of star-forming systems on smaller scales. It is questionable, if a simplified assumption about the structure of the system will lead to a reasonable constraint on the chemical evolution and the physical history of the system on all scales. Even the earliest phases of protostellar systems, Class 0 protostars, prior to the emergence of vivid protoplanetary discs, show evidence for multiple components. The CALYPSO survey targeted 16 Class 0 protostars, of which only 2 displayed Keplerian rotation, but for 11 the dust continuum was better reproduced by a model with a disc-like component (\citealt{2014A&A...563L...1M,2020A&A...635A..15M}).  These results are further supported by investigation of the specific angular momentum profiles in the inner envelope of the targets (\citealt{2020arXiv200110004G}). ALMA data from recent years complicates this picture further, as it suggests that the structures of protostellar sources are significantly more complicated than assumed by 1D and 2D models (e.g., \citealt{2020arXiv200511954M}). Efforts in modelling the chemistry with 3D magnetohydrodynamics (MHD) simulations (e.g., \citealt{2016ApJ...822...12H,2018MNRAS.478.2723Z}) also need to be considered. Thus, 1D models considering infall pathways purely within a cold, dense envelope are unlikely to produce representative chemical abundances in hot corinos surrounding protostars, even if some observational constraints are applied.
\newline 
\noindent This should not lessen the importance of improving and testing the chemical networks. Included reactions rates need to be carefully evaluated and in many cases, data or its accuracy are lacking (for example, even for the well-studied hydrogenation of CO differences exist in the investigated models, Section \ref{sec_postcoll_redone}). Furthermore, chemical networks are limited by computational capacities, which induces potential bottlenecks. If this is then coupled to a simplified assumption in the physical model, it is impossible to allocate from where differences between theoretical predictions and observational data arise. However, if no data of a source of interest is available to place constraints on the physical parameters, a simplified assumption is a valid approach to better constrain the physical parameters that can play a role in the formation or destruction of some species. Nonetheless, including multiple components in the physical model, whether derived from observations for specific sources or in a generic approach to predict the chemical evolution over a physical parameter range should improve our understanding of crucial chemical processes during the various stages of star formation.

\section{Conclusions}\label{Conclusion}
This work investigates different physicochemical modelling approaches to low-mass star formation. Benefits and caveats of using a simplified 1D or a more generic 2D physical model alongside two-phase or three-phase chemical networks have been discussed based on abundant, simple molecular species. While these models are powerful tools to derive the chemical evolution of these systems and investigate its initial formation conditions, seemingly small deviations of the input parameters can lead to significant differences in the output over astrophysical timescales. The major findings are raised as follows:
\begin{itemize}
    \item Assumptions about the precollapse phase need to be drawn carefully. Small differences in the physical parameters have shown to lead to significant deviations if evolved over the lifetime of a core. Instead of adopting standard literature values, a wider range of physical parameters and the consequent range of abundances should be taken into account.
    \item The precollapse duration impacts not only the initial molecular budget at the beginning of the collapse (i.e., that of a prestellar core), but also the subsequent chemical evolution of the star-forming system. Even within applied constraints from observations of pre- and protostellar systems, the deduced results vary significantly. Disagreements are exacerbated if the initial elemental abundances vary by more than an order of magnitude.
    \item The more advanced two-phase approach to modelling the surface chemistry does not impact the presented results much. For simple molecular species, no strong differences in abundances arise from the implementation of a two- or three-phase model for similar physical environments. Observations can not differentiate between surface and bulk ice, which makes the accurate representation of the solid phase under the diverse interstellar conditions challenging in models.
    \item A simplified physical model can be suitable for a first estimate to draw conclusions based on observations of low spatial resolution, when the underlying structures are missed. However, deviations between physicochemical models and observations should not always be attributed to caveats in the chemical network by default. Whether source-tailored or not, mismatch may stem from the assumption of a more simplified physical structure.
    \item High spatial resolution observations of inner regions of low-mass star-forming systems that probe the disc regime must be coupled with more sophisticated models that represent the relevant structures. The chemical composition of such systems can be computed accurately only if the duration of the collapse and the timing of protostellar birth can be constrained. Furthermore, radiative transfer molecular line simulations of contracting dense cores need to be carried out to place constraints on the dynamical evolution and time scales relevant for chemistry. 
\end{itemize}
Generic models with more comprehensive physics may not provide the optimal match to observations, but allow a source to be studied in perspective of other star-forming regions. Future observations on even small scales will require more detailed physicochemical models.

\section*{Acknowledgements}
The authors thank the referee P. Caselli for her helpful comments and suggestions that improved the paper. BMK and MND acknowledge the Swiss National Science Foundation (SNSF) Ambizione grant 180079. MND also acknowledges the Center for Space and Habitability (CSH) Fellowship, and the IAU Gruber Foundation Fellowship. AC acknowledges financial support from the Agence Nationale de la Recherche (grant ANR-19-ERC7-0001-01). SM acknowledges support from the H2020 European Research Council (ERC) (grant agreement No 646908) through ERC
Consolidator Grant ``S4F''. Research at Centre for Star and Planet Formation is funded by the Danish National Research Foundation. GS thanks T. Hosokawa for providing his stellar model results in digital form. GS also thanks R. Garrod for support, and the department of chemistry at UVA for funding the project.

\section*{Data Availability}
The data underlying this article are available in the article.

\FloatBarrier

\bibliographystyle{mnras}
\bibliography{references.bib}

\FloatBarrier

\appendix

\section{Initial abundances}
\begin{figure}
    \centering
    \includegraphics[width=1.0\columnwidth]{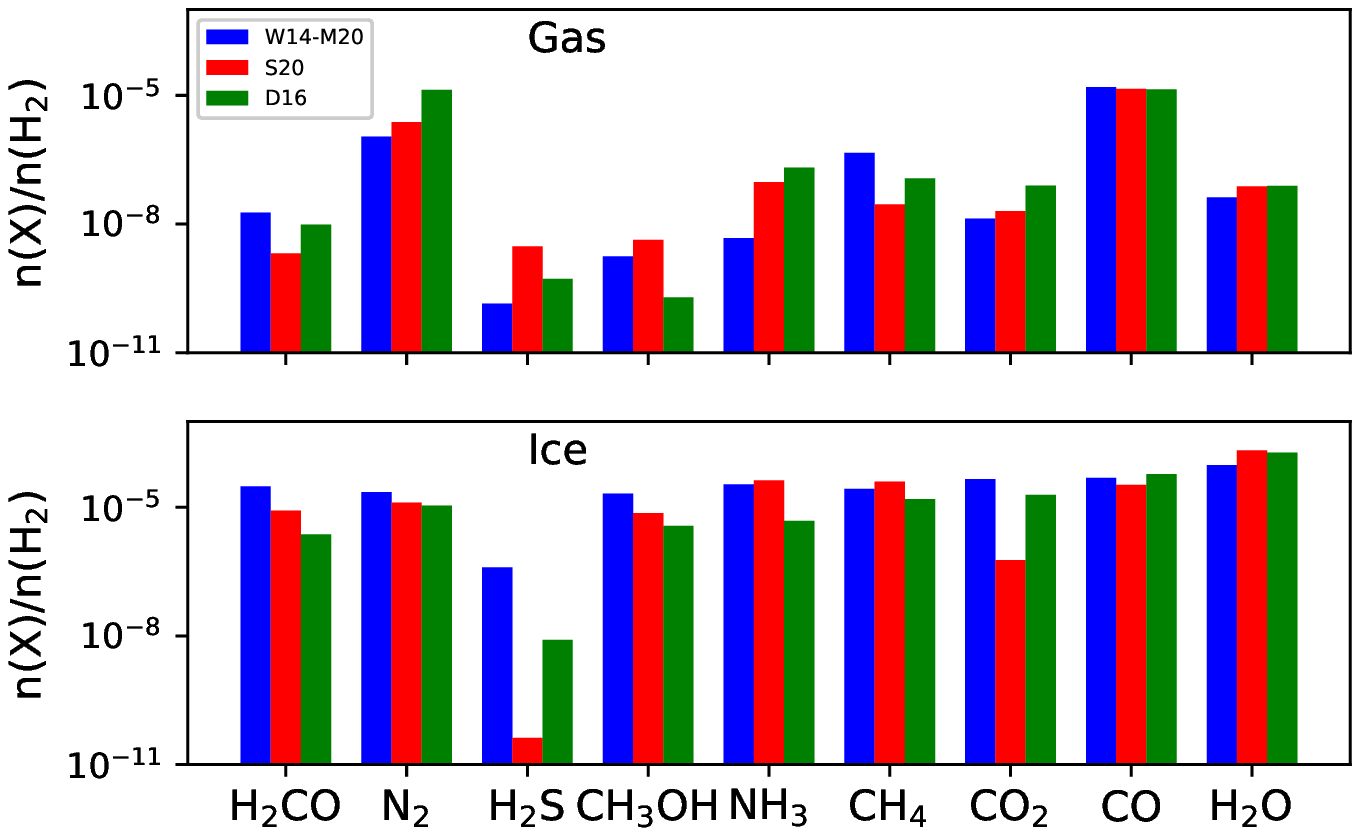}
    \caption{Molecular precollapse abundances (Table \ref{initial_molecular}) at the start of the collapse phase relative to n$_{\rm H_{2}}$. Blue bars describe the values of the W14-M20 model, red bars describe the S20 model, and green bars describe the D16 model. The upper panel depicts the gases, the lower panel shows the ices. For the three-phase chemical models (\textsc{Nautilus} and \textsc{Magickal}, used in W14-M20 and S20, respectively), the bulk and surface have been summed to obtain the total abundances in the ice.}
    \label{initialmol}
\end{figure}

This section provides additional information about the abundances before and after the prestellar phase. In Table \ref{initial_abundances}, the atomic abundances relative to n$_{\rm H}$ at the beginning of the precollapse phase are presented. These are used to compute the initial molecular abundances that are used as input for the chemical modelling of the collapse. These abundances are given in Table \ref{initial_molecular} and are plotted for the gas and the ice in Fig. \ref{initialmol}.
\begin{table}
    \centering
    \caption{Initial atomic abundances at the start of the precollapse phase of the three studied models relative to n$_{\rm H}$.}
    \begin{tabular}{|l|l|l|l|}
    \hline 
         & W14-M20$^{\rm a}$ & S20$^{\rm b}$ & D16$^{\rm c}$  \\
         \hline 
        H & -  & 1.0~$\times$~10$^{-4}$ & 5.0~$\times$~10$^{-5}$ \\
        H$_2$ & 0.5 & 0.49995 & 0.5  \\
        He & 9.0~$\times$~10$^{-2}$ & 9.0~$\times$~10$^{-2}$ & 9.8~$\times$~10$^{-2}$  \\
        C$^{\rm d}$ & 1.7~$\times$~10$^{-4}$ & 1.4~$\times$~10$^{-4}$ & 1.4~$\times$~10$^{-4}$ \\
        N & 6.2~$\times$~10$^{-5}$ & 7.0~$\times$~10$^{-5}$ & 7.5~$\times$~10$^{-5}$ \\
        O & 2.4~$\times$~10${-4}$ & 3.2~$\times$~10$^{-4}$ & 3.2~$\times$~10$^{-4}$ \\
        Na & - & 2.0~$\times$~10$^{-8}$ & 2.0~$\times$~10$^{-9}$ \\
        Mg & - & 7.0~$\times$~10$^{-9}$ & 7.0~$\times$~10$^{-9}$ \\
        Si & - & 8.0~$\times$~10$^{-9}$ & 8.0~$\times$~10$^{-9}$ \\
        P & - & 3.0~$\times$~10$^{-9}$ & 3.0~$\times$~10$^{-9}$ \\
        S$^{\rm d}$ & 1.5~$\times$~10$^{-6}$ & 8.0~$\times$~10$^{-8}$ & 8.0~$\times$~10$^{8}$ \\
        Cl$^{\rm d}$ & 1.0~$\times$~10$^{-9}$ & 4.0~$\times$~10$^{-9}$ & 4.0~$\times$~10$^{-9}$ \\
        Fe$^{\rm d}$ & 1.0~$\times$~10$^{-8}$ & 3.0~$\times$~10$^{-9}$ & 3.0~$\times$~10$^{-9}$ \\
        F & 6.68~$\times$~10$^{-9}$ & - & - \\
    \hline 
    \end{tabular}
    \begin{tablenotes}
    \item[*] $^{\rm a}$ physical model from \cite{2014MNRAS.445.2854W}, based on \cite{2008ApJ...674..984A}, last published in \cite{2020arXiv200704000M}; $^{\rm b}$  St\'{e}phan et al. (2020, in prep.); $^{\rm c}$ last published in \cite{2016MNRAS.462..977D}; $^{\rm d}$ in the case of W14-M20 these elements are fully in their ionic form (C$^+$, S$^+$, Cl$^+$, and Fe$^+$) 
    \end{tablenotes}
    \label{initial_abundances}
\end{table}

\begin{table}
    \centering
    \caption{Precollapse molecular abundances for the three studied models relative to n$_{\rm H_{2}}$ at the start of the collapse phase.}
    \begin{tabular}{|l|l|l|l|}
    \hline 
         & W14-M20$^{\rm a}$  & S20$^{\rm b}$  & D16$^{\rm c}$  \\
    \hline 
    gas  & & &  \\
    \hline 
        H$_{2}$CO & 1.85~$\times$~10$^{-8}$ & 2.09~$\times$~10$^{-9}$  & 7.73~$\times$~10$^{-8}$ \\
        N$_2$ & 1.10~$\times$~10$^{-6}$ & 2.38~$\times$~10$^{-6}$ & 1.35~$\times$~10$^{-5}$ \\
        H$_2$S & 1.39~$\times$~10$^{-10}$ & 3.01~$\times$~10$^{-9}$ & 5.28~$\times$~10$^{-10}$   \\
        CH$_3$OH & 1.77~$\times$~10$^{-9}$ & 4.26~$\times$~10$^{-9}$ & 1.97~$\times$~10$^{-10}$   \\
        NH$_3$ & 4.71~$\times$~10$^{-9}$ & 9.52  $\times$  10$^{-8}$  & 2.08~$\times$~10$^{-7}$  \\
        CH$_4$ & 4.56~$\times$~10$^{-7}$ & 2.86~$\times$~10$^{-8}$ & 1.16~$\times$~10$^{-7}$   \\
        CO$_2$ & 1.34~$\times$~10$^{-8}$ & 2.00~$\times$~10$^{-8}$& 7.89~$\times$~10$^{-8}$   \\
        CO & 1.56~$\times$~10$^{-5}$ & 1.41~$\times$~10$^{-5}$& 1.39~$\times$~10$^{-5}$   \\
        H$_2$O & 4.19~$\times$~10$^{-8}$ & 7.58~$\times$~10$^{-8}$ & 7.73~$\times$~10$^{-8}$  \\
    \hline 
     ice & & & \\
    \hline
    H$_{2}$CO & 3.07~$\times$~10$^{-5}$ & 8.31~$\times$~10$^{-6}$ & 2.32~$\times$~10$^{-6}$ \\
    N$_2$ & 2.26~$\times$~10$^{-5}$ & 1.27~$\times$~10$^{-5}$ & 1.09~$\times$~10$^{-5}$  \\
    H$_2$S & 3.93~$\times$~10$^{-7}$ &  4.21~$\times$~10$^{-11}$ & 8.11~$\times$~10$^{-9}$  \\
    CH$_3$OH & 2.08~$\times$~10$^{-5}$ & 7.80~$\times$~10$^{-6}$ & 3.66~$\times$~10$^{-6}$  \\
    NH$_3$ & 3.41~$\times$~10$^{-5}$ & 4.25~$\times$~10$^{-5}$ & 4.82~$\times$~10$^{-6}$  \\
    CH$_4$ & 2.67~$\times$~10$^{-5}$ & 3.97~$\times$~10$^{-5}$ & 1.54~$\times$~10$^{-5}$  \\
    CO$_2$ & 4.53~$\times$~10$^{-5}$ & 5.86~$\times$~10$^{-7}$ & 5.92~$\times$~10$^{-5}$  \\
    CO & 4.85~$\times$~10$^{-5}$ & 3.32~$\times$~10$^{-5}$ & 1.39~$\times$~10$^{-5}$  \\
    H$_2$O & 9.53~$\times$~10$^{-5}$ & 2.12~$\times$~10$^{-4}$ & 1.87~$\times$~10$^{-4}$  \\
    \hline 
    \end{tabular}
    \begin{tablenotes}
    \item[*] $^{\rm a}$ physical model from \cite{2014MNRAS.445.2854W}, based on \cite{2008ApJ...674..984A}, last published in \cite{2020arXiv200704000M}; $^{\rm b}$  St\'{e}phan et al. (2020, in prep.); $^{\rm c}$ last published in \cite{2016MNRAS.462..977D}  
    \end{tablenotes}
    \label{initial_molecular}
\end{table}

\FloatBarrier

\section{Individual abundances}
Here, the evolution of the abundances of the individual molecular species in the gas and the ice are presented.  Furthermore, an updated version of Fig. \ref{env_vs_disc} with results obtained with the two-phase chemical network is depicted.

\begin{figure}
    \centering
    \includegraphics[width=1.0\columnwidth]{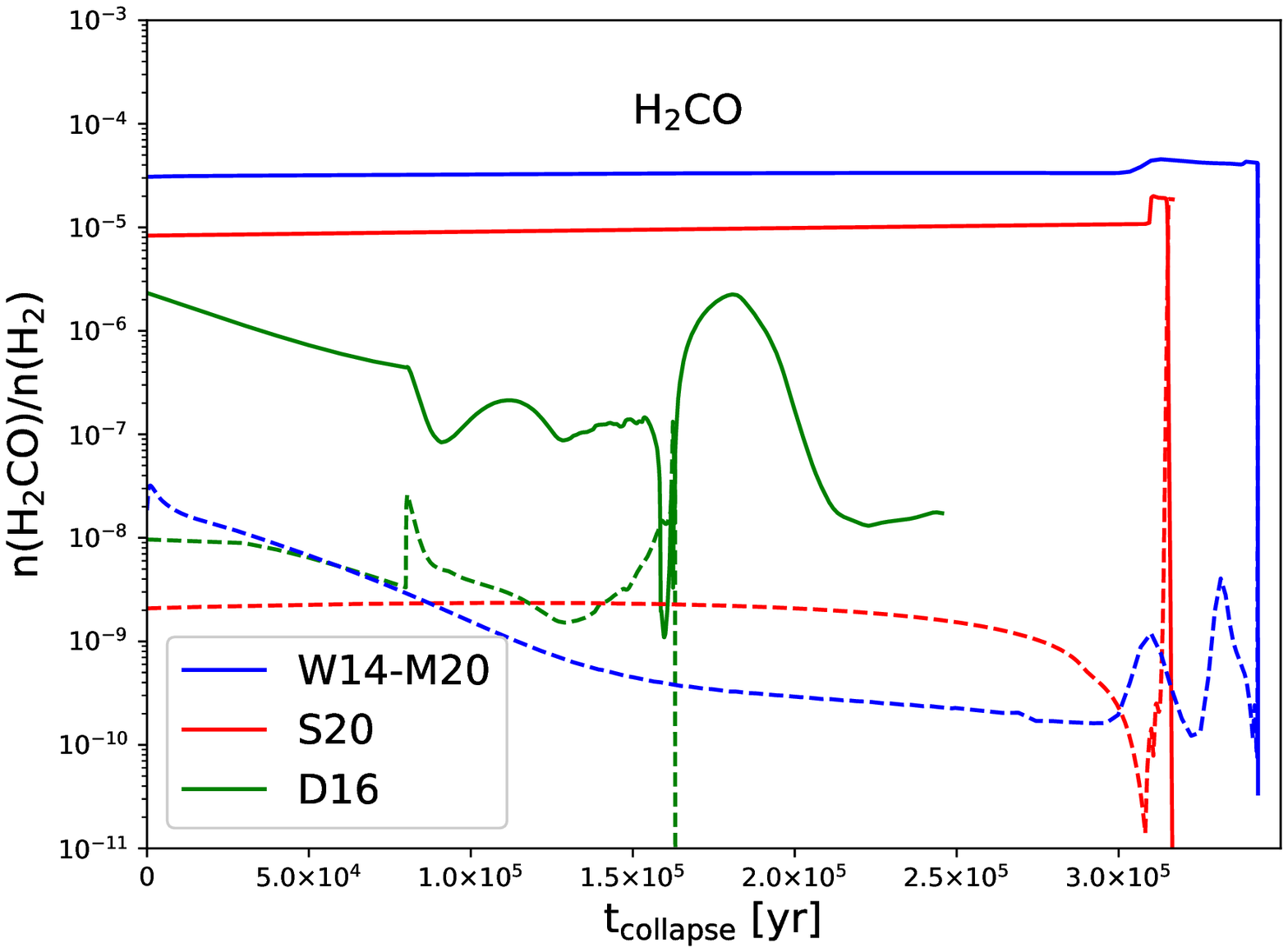}
    \caption{This figure shows the evolution of H$_2$CO throughout the collapse. The collapse time in years is plotted on the x-axis, the abundance relative to n$_{\rm H_{2}}$ is plotted on the y-axis. Dashed lines represent H$_2$CO$_{\rm gas}$, solid lines H$_2$CO$_{\rm ice}$. The trajectory of 62.4~au from the W14-M20 model is depicted in blue, the 49.7~au from S20, in red and the 46.7~au from D16 in green.  For the three-phase chemical models (\textsc{Nautilus} and \textsc{Magickal}, used in W14-M20 and S20, respectively), the bulk and surface have been summed to obtain the total abundances in the ice.}
    \label{H2CO}
\end{figure}

\begin{figure}
    \centering
    \includegraphics[width=1.0\columnwidth]{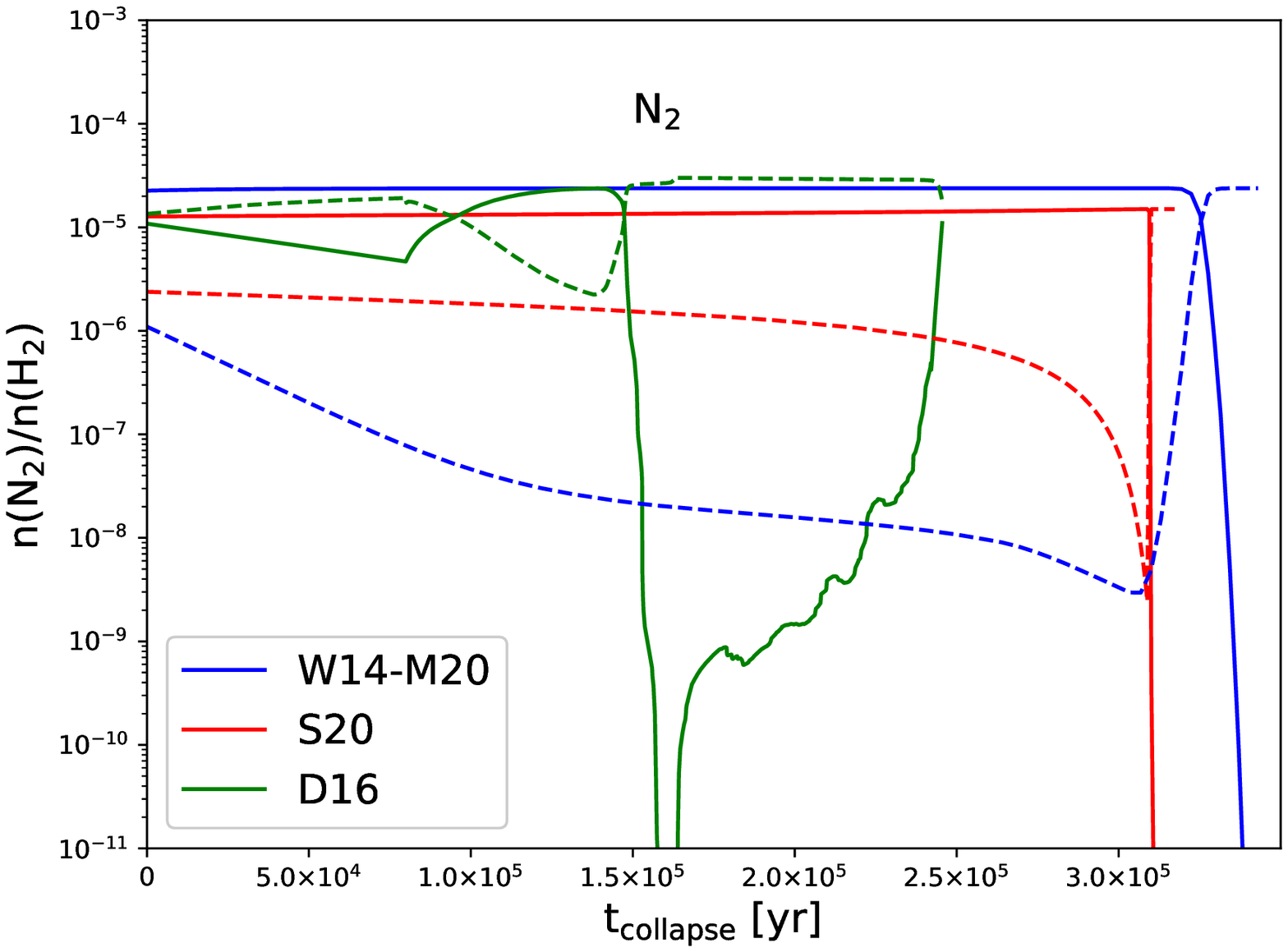}
    \caption{Same as Fig. \ref{H2CO}, but for N$_2$.}
    \label{N2}
\end{figure}

\begin{figure}
    \centering
    \includegraphics[width=1.0\columnwidth]{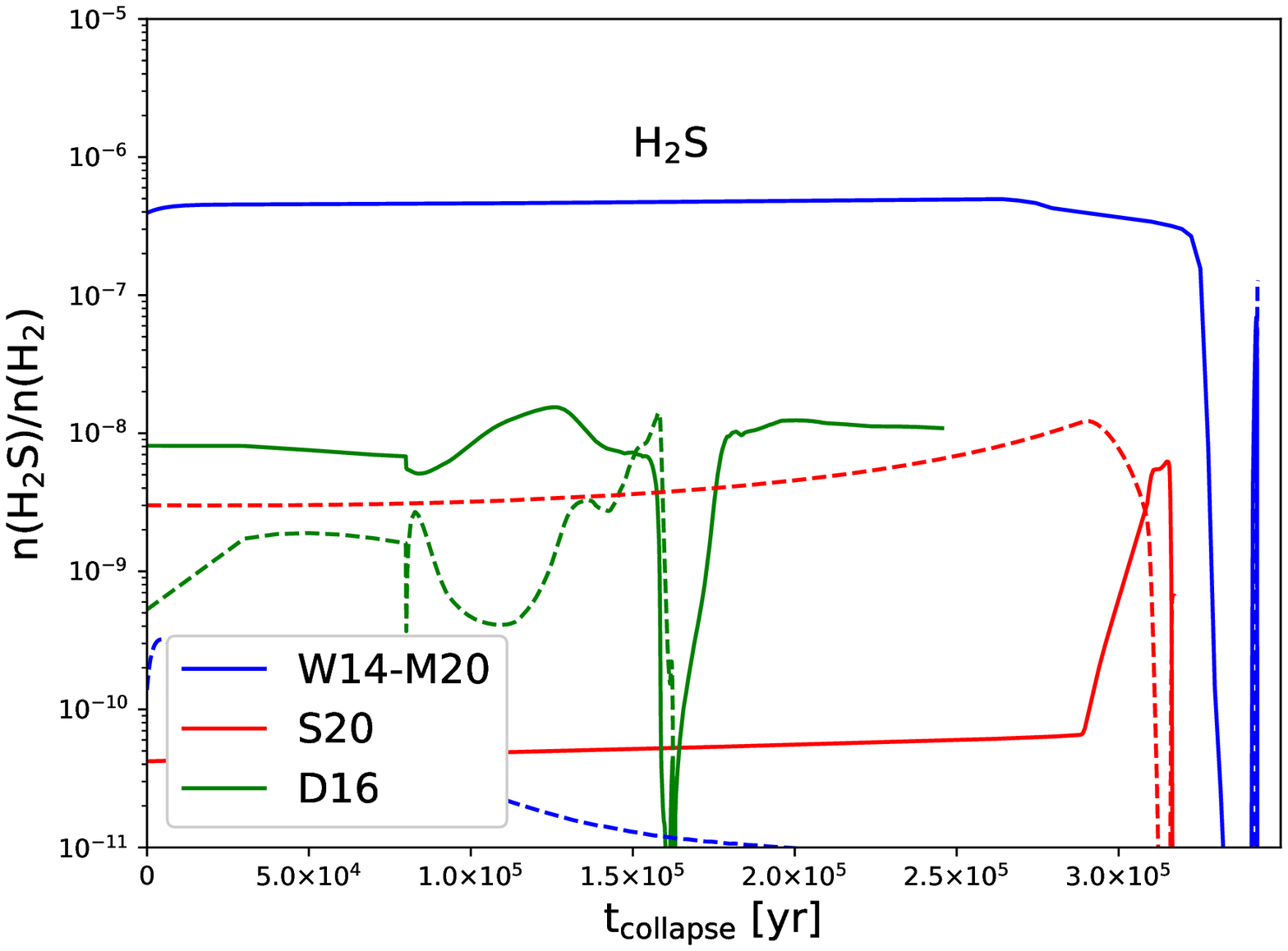}
    \caption{Same as Fig. \ref{H2CO}, but for H$_2$S.}
    \label{H2S}
\end{figure}

\begin{figure}
    \centering
    \includegraphics[width=1.0\columnwidth]{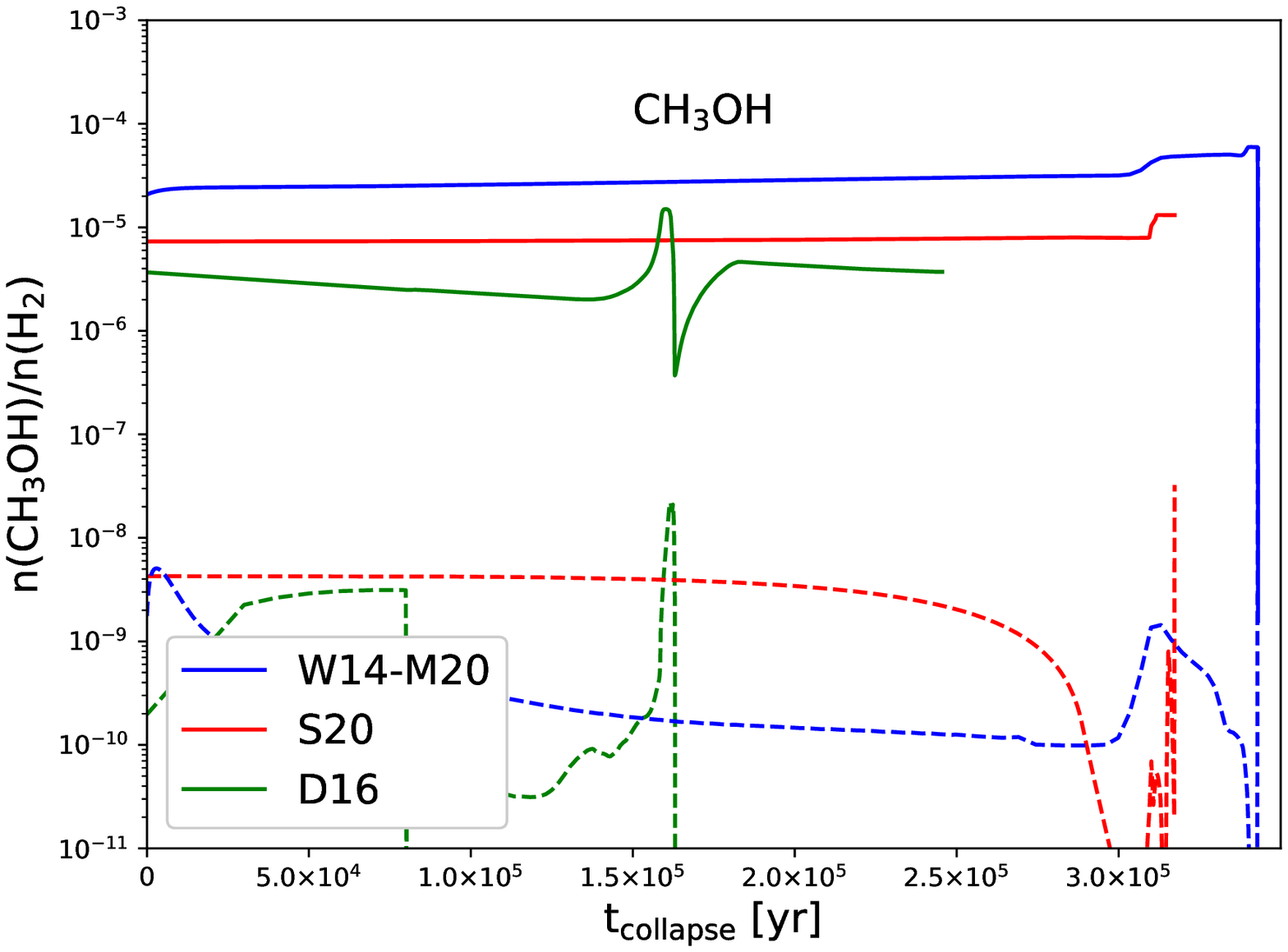}
    \caption{Same as Fig. \ref{H2CO}, but for CH$_3$OH.}
    \label{CH3OH}
\end{figure}

\begin{figure}
    \centering
    \includegraphics[width=1.0\columnwidth]{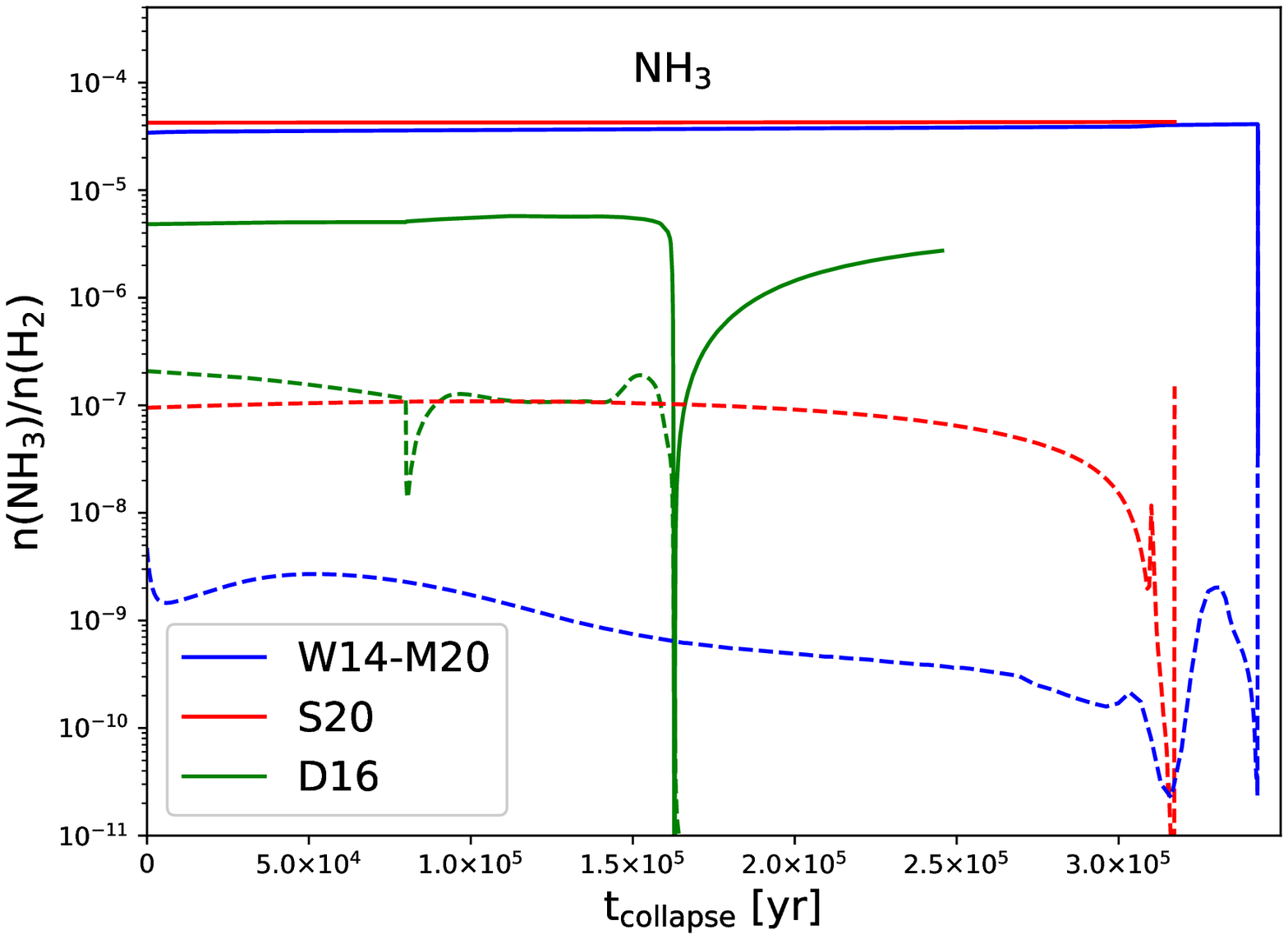}
    \caption{Same as Fig. \ref{H2CO}, but for NH$_3$.}
    \label{NH3}
\end{figure}

\begin{figure}
    \centering
    \includegraphics[width=1.0\columnwidth]{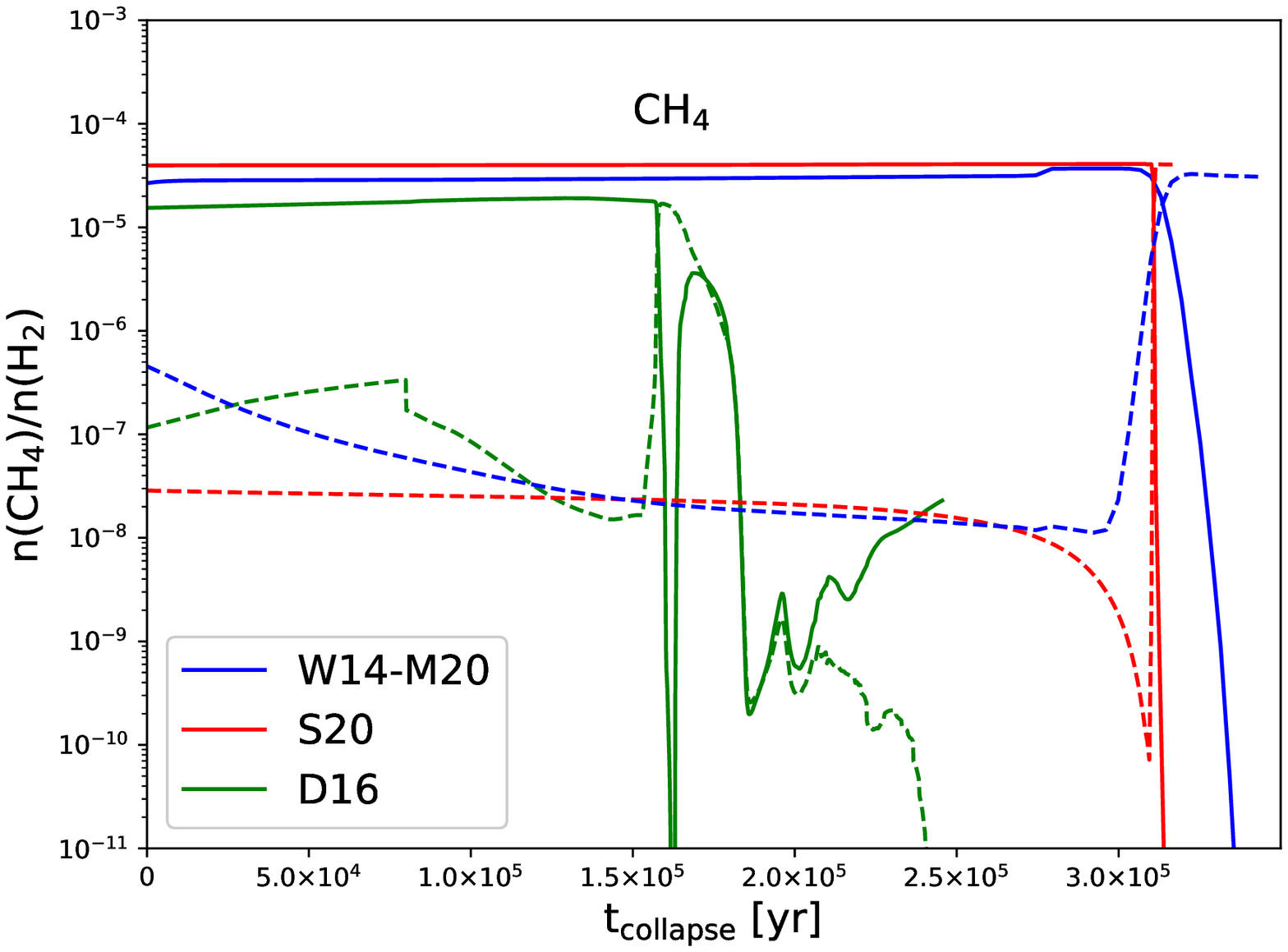}
    \caption{Same as Fig. \ref{H2CO}, but for CH$_4$.}
    \label{CH4}
\end{figure}

\begin{figure}
    \centering
    \includegraphics[width=1.0\columnwidth]{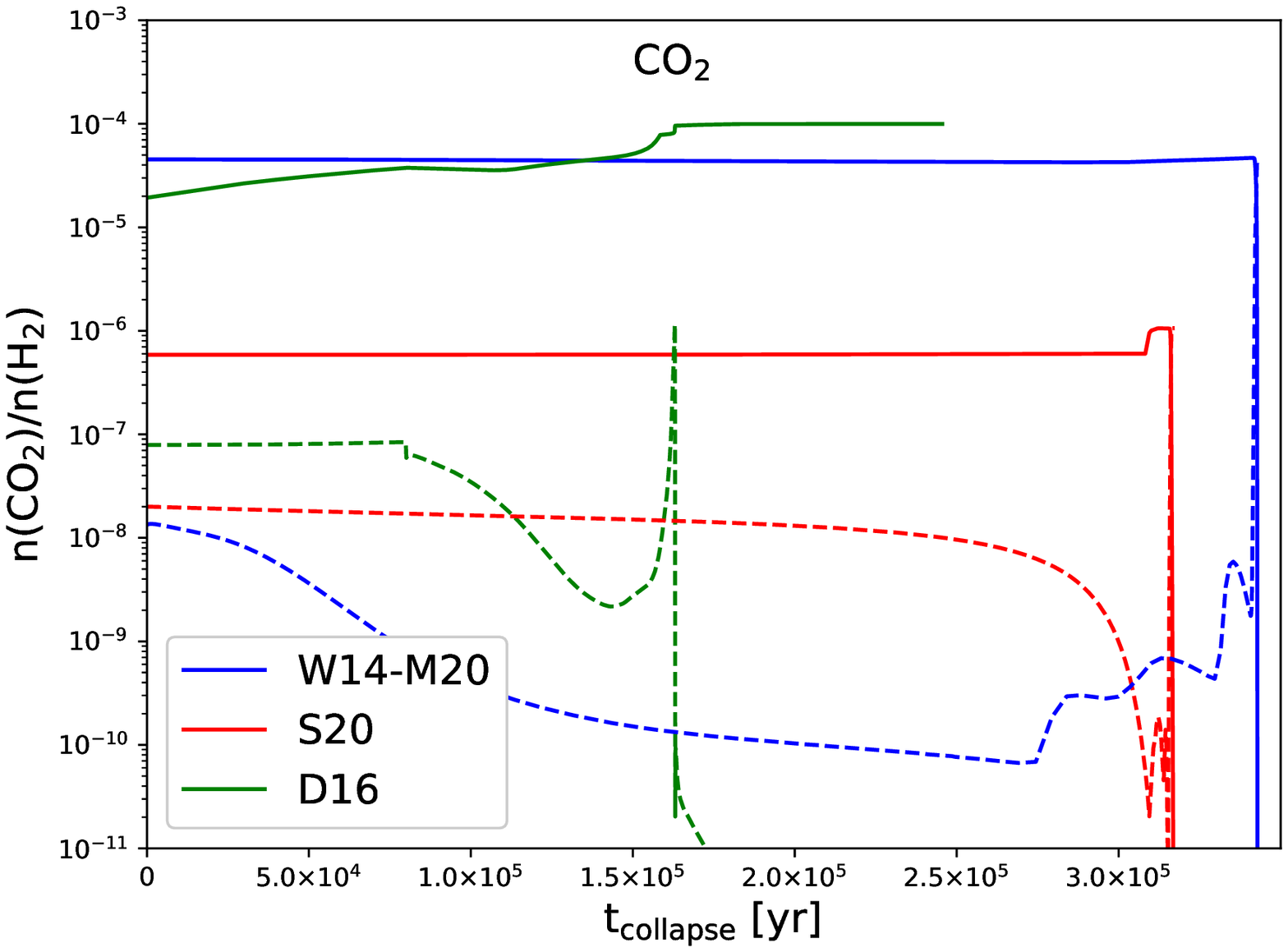}
    \caption{Same as Fig. \ref{H2CO}, but for CO$_2$.}
    \label{CO2}
\end{figure}

\begin{figure}
    \centering
    \includegraphics[width=1.0\columnwidth]{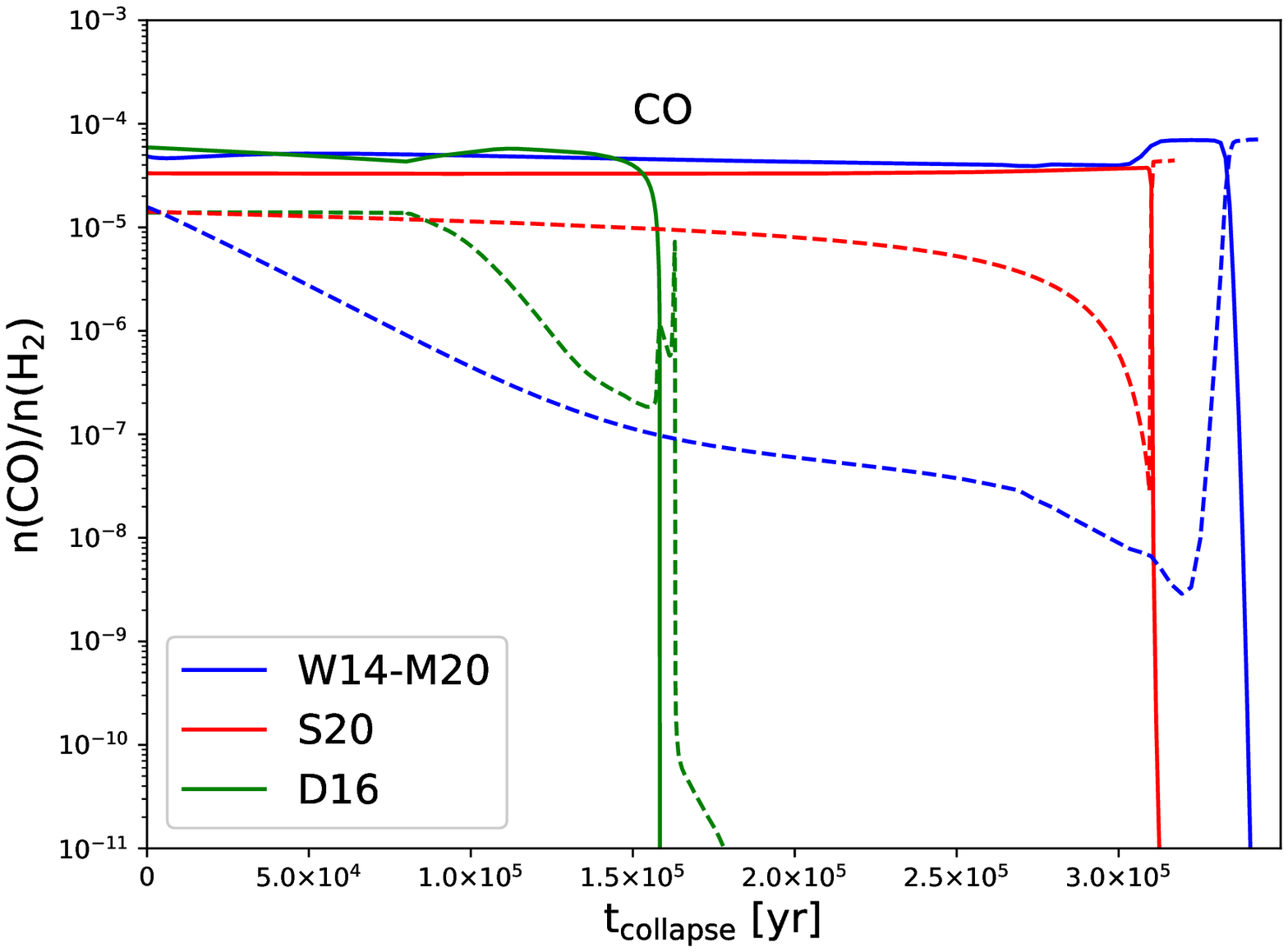}
    \caption{Same as Fig. \ref{H2CO}, but for CO.}
    \label{CO}
\end{figure}

\begin{figure}
    \centering
    \includegraphics[width=1.0\columnwidth]{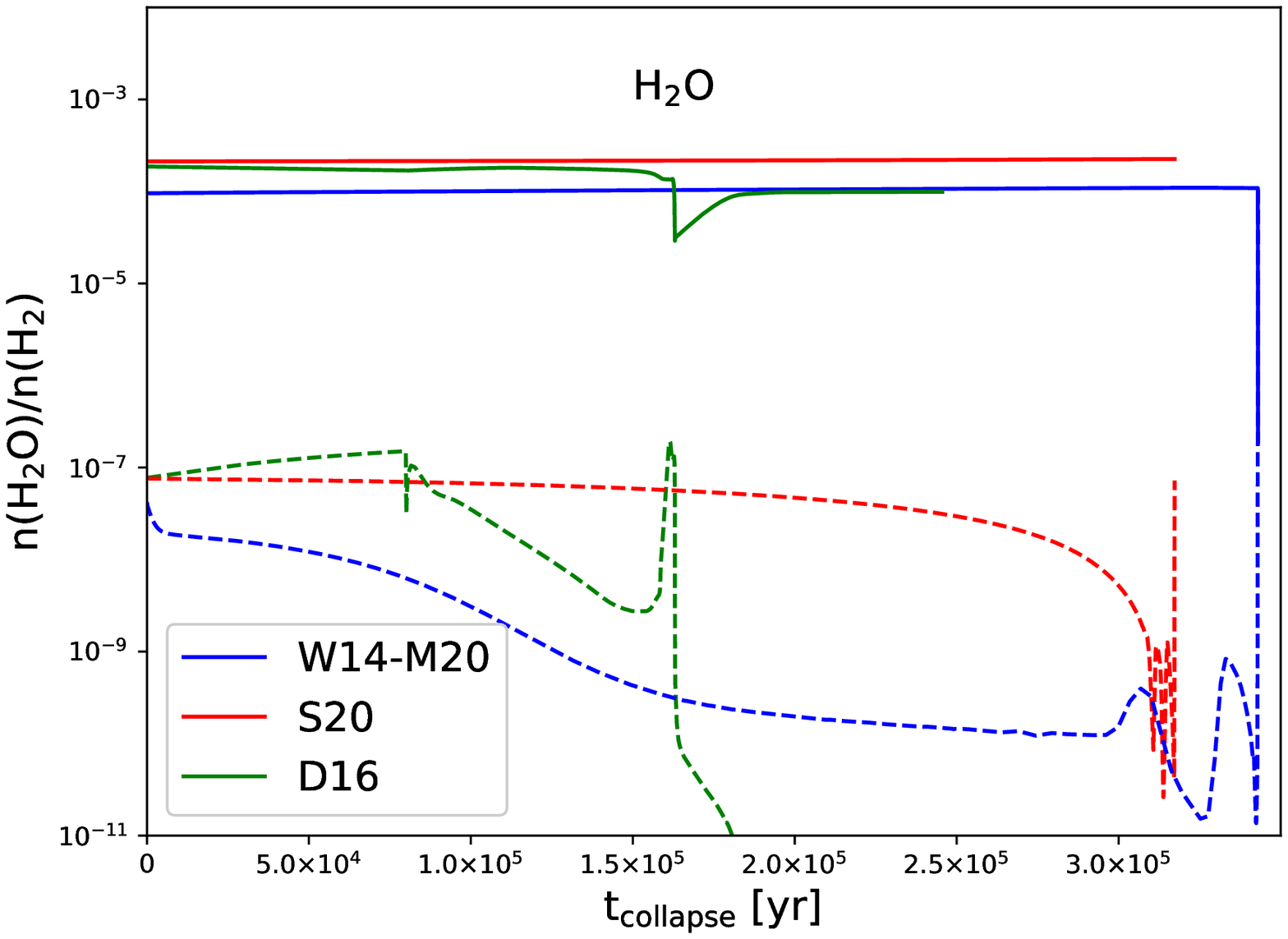}
    \caption{Same as Fig. \ref{H2CO}, but for H$_2$O.}
    \label{H2O}
\end{figure}

\begin{figure}
    \centering
    \includegraphics[width=1.0\columnwidth]{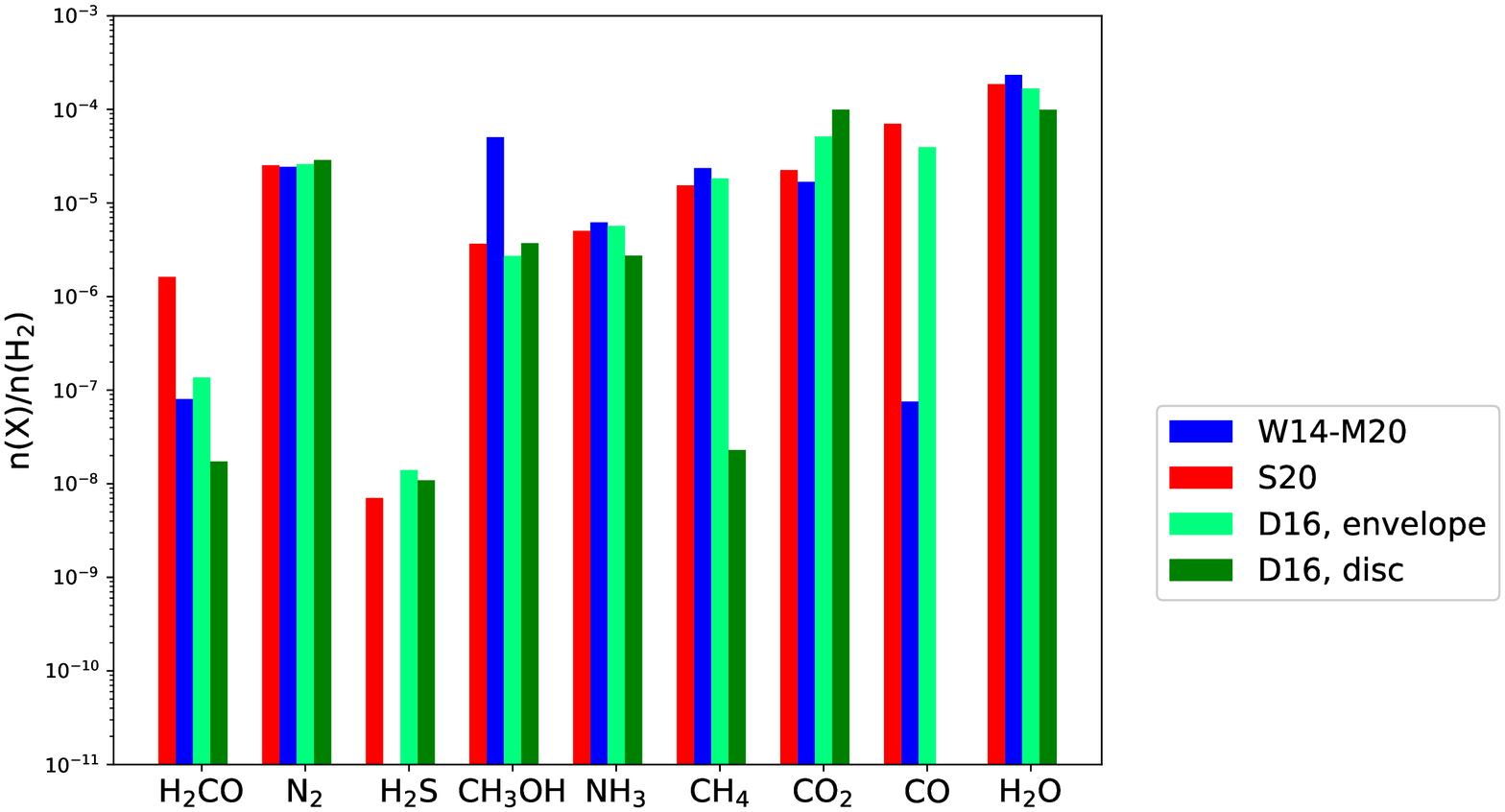}
    \caption{Post-collapse molecular abundances relative to n$_{\rm H_{2}}$ with the two-phase D16 chemical model. Blue bars describe the values of the W14-M20 model (final position at 62.4 au), red bars show the S20 model (final position at 49.7 au), light green bars correspond to the abundances in the warm envelope upon disc entry in the D16 model (at t~=~1.5~$\times$~10$^5$~yr =~0.61~t$_{\rm collapse}^{\rm D16}$), dark green bars illustrate the abundances in the disc at the end of the collapse in the D16 model (final position at 46.7 au).}
    \label{env_vs_disc_newchem}
\end{figure}

\bsp	
\label{lastpage}
\end{document}